\documentclass[12pt]{article}

\usepackage{scicite}
\usepackage{graphicx}
\usepackage{times}
\usepackage{siunitx}
\usepackage{longtable}
\usepackage{url}
\usepackage{lineno}
\usepackage{amssymb}
\usepackage[T5,T1]{fontenc} 
\usepackage{placeins}

\newenvironment{sciabstract}{%
\begin{quote} \bf}
{\end{quote}}

\title{Observation of high-energy neutrinos from the Galactic plane}

\author
{IceCube Collaboration*\textsuperscript{\textdagger}\\
\\
\normalsize{*E-mail: analysis@icecube.wisc.edu. \textsuperscript{\textdagger} IceCube Collaboration authors and affiliations}\\
\normalsize{are listed in the supplementary material.}\\
}

\date{}

\begin{document} 


\baselineskip24pt


\maketitle


\begin{sciabstract}

The origin of high-energy cosmic rays, atomic nuclei that continuously impact Earth's atmosphere, has been a mystery for over a century. Due to deflection in interstellar magnetic fields, cosmic rays from the Milky Way arrive at Earth from random directions.  However, near their sources and during propagation, cosmic rays interact with matter and produce high-energy neutrinos. We search for neutrino emission using machine learning techniques applied to ten years of data from the IceCube Neutrino Observatory. We identify neutrino emission from the Galactic plane at the 4.5$\sigma$ level of significance, by comparing diffuse emission models to a background-only hypothesis. The signal is consistent with modeled diffuse emission from the Galactic plane, but could also arise from a population of unresolved point sources.

\end{sciabstract}



\begin{figure}
    \centering
    \includegraphics[width=\textwidth]{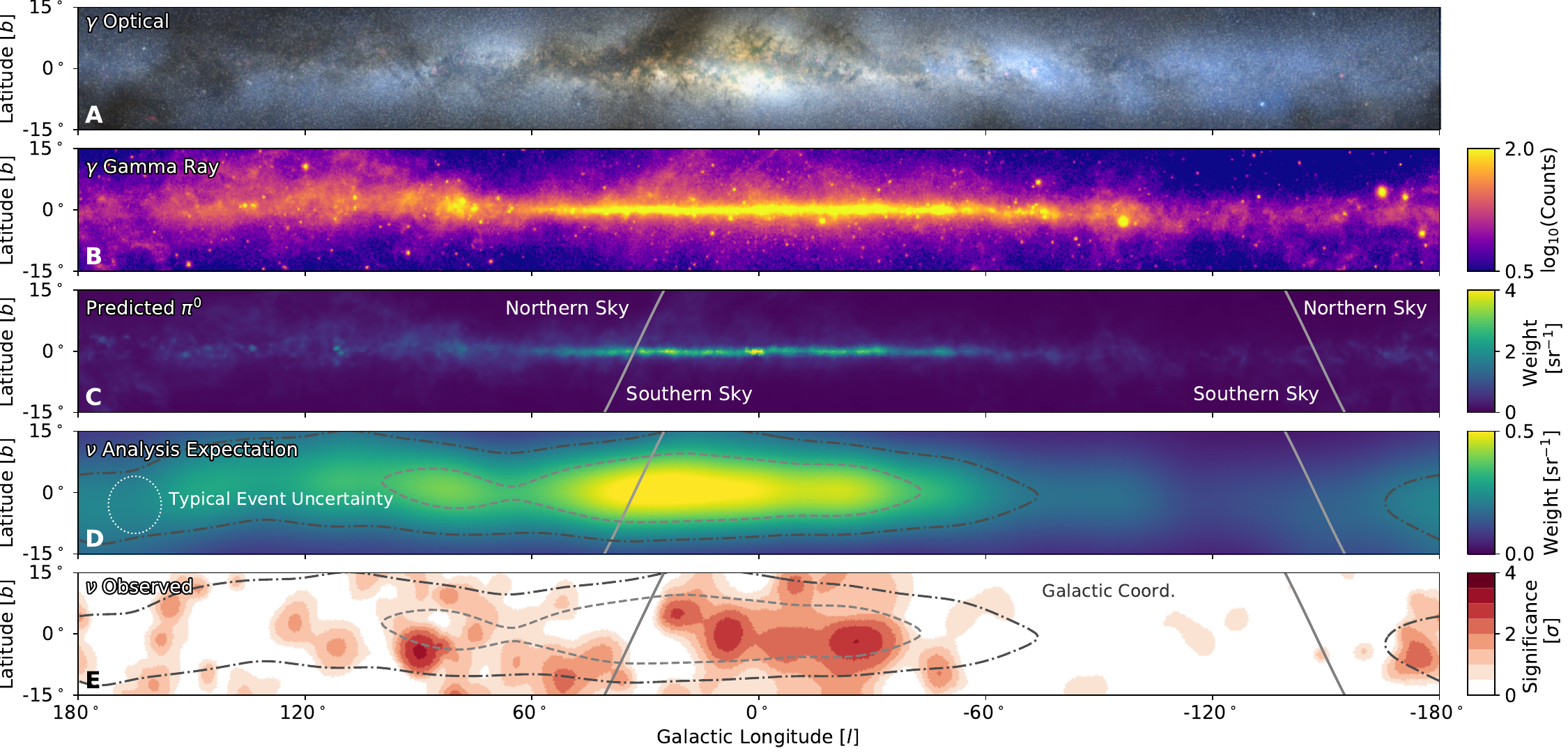}
    \caption{\textbf{The plane of the Milky Way galaxy in photons and neutrinos.} Each panel is in Galactic coordinates, with the origin being at the Galactic Center, extending $\pm15^\circ$ in latitude and $\pm180^\circ$ in longitude. (A) Optical color image~\cite{optical}, which is partly obscured by clouds of gas and dust that absorb optical photons. Credit A. Mellinger, used with permission. (B) The integrated flux in gamma rays from the  {\it Fermi} Large Area Telescope ({\it Fermi}-LAT) 12 year survey~\cite{12yr-gammamap} at energies greater than 1\,GeV, obtained from the {\it Fermi} Science Support Center and processed with the {\it Fermi}-LAT ScienceTools. (C) The emission template calculated for the expected neutrino flux, derived from the $\pi^0$ template that matches the {\it Fermi}-LAT observations of the diffuse gamma-ray emission~\cite{fermi}.  (D) The emission template from panel (C) including the detector sensitivity to cascade-like neutrino events and the angular uncertainty of a typical signal event (7$^\circ$, indicated by the dotted white circle). Contours indicate the central regions that contain 20\% and 50\% of the predicted diffuse neutrino emission signal.  (E) The pre-trial significance of the IceCube neutrino observations, calculated from all-sky scan for point-like sources using the cascade neutrino event sample. Contours are the same as panel (D).  Grey lines in (C) - (E) indicate the Northern-Southern sky horizon line at the IceCube detector.}
    \label{fig:gp_inserts}
\end{figure}

The Milky Way emits radiation  across the electromagnetic spectrum, from radio waves to gamma rays.  Observations at different wavelengths provide insight into the structure of the Galaxy and have identified sources of the highest energy photons. For gamma rays with energies above 1 giga-electronvolt (GeV), the plane of the Milky Way is the most prominent feature visible on the sky (Figure~\ref{fig:gp_inserts}B). Most of this observed gamma-ray flux consists of photons generated by the decay of neutral pions ($\pi^0$), themselves produced by cosmic rays (high energy charged particles) colliding with the interstellar medium within the Milky Way galaxy~\cite{fermi}.

Photons can also be produced by interactions of energetic electrons with interstellar photon fields  or absorbed by ambient interstellar matter, so other messengers are needed to provide additional information on the cosmic-ray interactions and acceleration sites in the Galaxy. In addition to neutral pions, cosmic-ray interactions also produce charged pions which produce neutrinos when they decay.  Unlike photons, neutrinos rarely interact on their way to Earth, and so they directly trace the location and energetics of the cosmic-ray interactions.
As both gamma rays and neutrinos arise from the decay of pions, a diffuse neutrino flux concentrated along the Galactic plane has been predicted~\cite{Stecker1979,BEREZINSKY1993,Ingelman:1996md,Evoli:2007iy}.
Figure~\ref{fig:gp_inserts}D shows the expected tera-electronvolt-energy (TeV) neutrino flux, calculated from the GeV-energy {\it Fermi}-LAT observation~\cite{fermi}.
In addition to the predicted diffuse emission, the Milky Way is densely populated with numerous high-energy gamma-ray point sources (Figure~\ref{fig:gp_inserts}B),  several classes of which are potential cosmic-ray accelerators and therefore candidate neutrino sources \cite{Fermi-LAT:2013iui, Guetta:2002hv, Bednarek:2003cv, Gonzalez-Garcia:2009bev, Halzen:2016seh}.
This makes the Galactic plane an expected source of neutrinos.

Previous searches for this signal using the IceCube and ANTARES (Astronomy with a Neutrino Telescope and Abyss environmental RESearch) neutrino detectors \cite{gptracks, mesecascades, JointAntaresIceCube, ANTARESOnlyGP} have not found a statistically significant signal (p-values $\ge 0.02$).
The development of deep learning techniques in data science has produced new tools \cite{CNNPaper, Event-Generator, mirco_huennefeld_2022_7412035} that can identify a larger number of neutrino interactions in detector data, with improved angular resolution over earlier methods. Here, we apply these deep learning tools to IceCube data to search for evidence of neutrinos from the Galactic plane, including searches for diffuse neutrino emission and point source emission from catalogs of known sources of GeV gamma-rays.

\section*{Cascade events in IceCube}


The IceCube Neutrino Observatory~\cite{IceCubeDetector}, located at the South Pole, is designed to detect high energy ($\gtrsim 1$ TeV) astrophysical neutrinos and identify their sources.
The detector construction, which deployed instruments within a cubic kilometer of clear ice, was completed in 2011. 5160 digital optical modules (DOMs) were placed at depths between $\SI{1.5}{km}$ and $\SI{2.5}{km}$ below the surface of the Antarctic glacier.
Neutrinos are detected via Cherenkov radiation emitted by charged secondary particles that are produced by neutrino interactions with nuclei in the ice or bedrock.
Due to the large momentum transfer from the incoming neutrino, the directions of secondary particles are closely aligned with the incoming neutrino direction, enabling the identification of the neutrino's origin.
The two main detection channels are cascade and track events.
Cascades are short-ranged particle showers, predominantly from interactions of electron neutrino ($\nu_e$) and tau neutrino ($\nu_\tau$) with nuclei, as well as scattering interactions of all three neutrino flavors ($\nu_e$, muon neutrino ($\nu_\mu$) and $\nu_\tau$) on nuclei.
Because the charged particles in cascade events travel only a few meters,
these energy depositions appear almost point-like to IceCube's 
$\SI{125}{\meter}$ (horizontal) and 7\,m to 17\,m (vertical)
instrument spacing.  This results in larger directional uncertainties when compared to tracks, which are elongated, often several kilometers long, energy depositions arising predominantly from muons generated in cosmic-ray particle interactions in the atmosphere or muons created in interactions of $\nu_\mu$ with nuclei.
Unlike tracks, the energy deposited by cascades is often contained within the instrumented volume, providing a more complete measure of the neutrino energy~\cite{EnergyReconstructionPaper}.

Searches for astrophysical neutrino sources are affected by an overwhelming background of muons and neutrinos produced by cosmic-ray interactions with Earth's atmosphere.
Atmospheric muons dominate this background; IceCube records about 100 million muons for every observed astrophysical neutrino.
While muons from the Southern Hemisphere (above IceCube) can penetrate several kilometers deep into the ice,
muons from the Northern Hemisphere (below IceCube) are absorbed during passage through Earth.
Due to this shielding effect, and the superior angular resolution of tracks over cascades ($\lesssim 1^\circ$ compared to $\lesssim 10^\circ$ respectively, both above 10 TeV), searches for neutrino sources using IceCube typically rely on track selections, which are most sensitive to astrophysical sources in the Northern sky~\cite{10yrtracks}. 

However, the Galactic Center, as well as the bulk of the neutrino emission expected from the Galactic plane, is located in the Southern sky (Figure~\ref{fig:gp_inserts}C-D).
To overcome the muon background in the Southern sky, analyses of IceCube data often utilize events in which the neutrino interaction is observed within the detector~\cite{IceCube:2020wum,diffuse_cascades}.
The selection of these events greatly reduces the background rate of cosmic-ray muons which enter 
the instrumented volume from the Southern sky. Unlike these atmospheric muons, atmospheric neutrinos~\cite{Gaisser:2002jj}
generally cannot be distinguished by their interactions in the detector from neutrinos from astrophysical sources.
Nevertheless, with increasing energy, an increasing fraction of the atmospheric neutrinos from the Southern sky, above IceCube, can be excluded by eliminating events with simultaneous muons which originate from the same cosmic-ray air-shower that  produced the atmospheric neutrino~\cite{Self-Veto-1, Self-Veto}.
However, at TeV energies, relevant for searches of Galactic neutrino emission, the majority of these atmospheric neutrinos remain as a substantial background in searches for astrophysical neutrinos.
This background is dominated by muon neutrinos, which are largely detected as tracks in IceCube.
The selection of cascade events instead of track events therefore reduces the contamination of atmospheric neutrinos, by about an order of magnitude at TeV energies, and lowers the energy threshold of the analysis to about 1 TeV.

In the Southern sky, the lower background, better energy resolution, and lower energy threshold of cascade events compensate for their inferior angular resolution, compared to tracks.
This is particularly true for searches for emission from extended objects, such as the Galactic plane, for which the size of the emitting region is similar to (or larger than) the angular resolution.
Compared to track-based searches, 
cascade-based analyses are more reliant upon the signal purity and less on the angular resolution of individual events.
We therefore expect analyses based on cascades to have
substantially better sensitivity to extended neutrino emission in the TeV energy range from the Southern sky.

\section*{Application of deep learning to cascade events}

To identify and reconstruct cascade events in IceCube, we use tools based on deep learning.
These tools are designed to reject the overwhelming background from atmospheric muon events, then identify the energies and directions of the neutrinos that generated the cascade events.  
IceCube observes events at a rate of about about $\SI{2.7}{kHz}$~\cite{IceCubeDetector},
arising mostly from background events (atmospheric muons and atmospheric neutrinos) that outnumber signal events (astrophysical neutrinos) at a ratio of roughly $10^8$:1.
To search for neutrino sources, this event selection step is required to improve the signal purity by orders of magnitude.

Previously used event selections for cascade events~\cite{diffuse_cascades, MESE, HESE} relied on high-level observables, such as the  event location within the IceCube volume and total measured light levels,  to reduce the initial data rate.
In subsequent selection steps, more computing-intensive selection strategies were performed, such as the definition of veto regions within the detector, to further reject events identified as incoming muons.
We adopt a different approach, using  tools based on convolutional neural networks (CNNs)~\cite{ConvolutionalNetworks, CNNPaper} to perform event selections.  
The high inference speed of the neural networks (milliseconds per event) allows us to use a more complex filtering strategy at
earlier stages of the event selection pipeline.
This allows us to retain more lower energy astrophysical neutrino events (Figure~\ref{fig:effa}), and include cascade events that are difficult to reconstruct and distinguish from background due to their location at the boundaries of the instrumented volume or in regions of the ice with degraded optical clarity (due to higher concentrations of impurities in the ice).

After the selection of events, we refine event properties, such as the direction of the incoming neutrino and deposited energy, using the patterns of deposited light in the detector.
The likelihood of the observed light pattern under a given event hypothesis is maximized to determine the event properties that best describe the data.
For this purpose, a hybrid reconstruction method~\cite{Event-Generator, mirco_huennefeld_2022_7412035} is utilized that combines a maximum likelihood estimation with deep learning.
In this approach, a neural network (NN) is used to parameterize the relationship between the event hypothesis and expected light yield in the detector.
This smoothly approximates a more computationally-expensive Monte Carlo simulation, while avoiding the simplifications that limit other reconstruction methods~\cite{EnergyReconstructionPaper, IceCube:2021oqo}.
Starting with an event hypothesis, the NN models the photon yield at each DOM. Symmetries (e.g. rotation, translation, and time invariance of the neutrino interaction) and detector-specific domain knowledge are exploited by directly including them in the network architecture, analogous to how a Monte Carlo simulation would exploit this information.
This differs from previous CNN-based methods used in neutrino telescopes~\cite{CNNPaper}, which inferred the event properties directly from the observed data. 
However, the observed IceCube data is already convolved with detector effects, making it difficult to exploit the underlying symmetries.
Our hybrid method is intended to provide a more complete utilization of available information.
A description of the hybrid method has been published previously~\cite{Event-Generator} and we discuss its application to our dataset~\cite{note:supplmat}.

We find that this deep learning event selection retains more than 20 times as many events as the selection method used in the previous cascade-based Galactic plane analysis of IceCube data~\cite{mesecascades} (Figure~\ref{fig:effa}).  It also provides improved angular resolution, by up to a factor of 2 at TeV energies~\cite{Event-Generator} (Figure~\ref{fig:angular_resolution}).
The increased event rate is mostly due to the reduced energy threshold and the inclusion of events near the boundaries of the instrumented volume (Figure~\ref{fig:containment}). 
We analyzed ten years of IceCube data, collected between May 2011 and May 2021. A total of 59,592 events are selected over the entire sky in the 500\,GeV to multiple peta-electronvolts (PeV) energy range (compared to 1,980 events from seven years in the previous selection~\cite{mesecascades}).  
We estimate the remaining sample has an atmospheric muon contamination of about 6\%~\cite{note:supplmat}, while the astrophysical neutrino contribution is estimated to about 7\%, assuming the observed flux~\cite{diffuse_cascades}.
The remaining 87\% of the events are atmospheric neutrinos.
These fractions are not used in the analysis directly; instead the entire sample is used to derive a data-driven background estimate.

\section*{Searches for Galactic neutrino emission}


We used this event selection to perform searches based on several neutrino emission hypotheses~\cite{note:supplmat}. For each hypothesis, we use a previously-described maximum-likelihood-based method~\cite{llhmethod}, modified to account for signal contamination in the data-derived background model~\cite{gptracks, mesecascades}. 
These techniques, decided a priori and blind to the reconstructed event directions,  infer the background from the data itself, avoiding the uncertainties introduced by background modelling. 
P-values are calculated by comparing the experimental results with mock experiments performed on randomized experimental data. 
The backgrounds for these searches, consisting of atmospheric muons, atmospheric neutrinos, and the flux of extra-galactic astrophysical neutrinos, are each largely isotropic. The rotation of  Earth ensures that, for a detector located at the South Pole, the detector sensitivity to neutrinos in right ascension is fairly uniform in each declination band. Therefore, we estimate backgrounds by scrambling the right ascension value of each event, preserving all detector-specific artifacts in the data.
Any systematic differences between the modeling of signal hypotheses and the true signal could reduce the 
sensitivity of our search, but do not change the resulting p-values.
 
The source hypothesis tests  were defined a priori.  They include tests for the diffuse emission expected from cosmic rays interacting with the interstellar medium in the Galactic plane, tests using catalogs of known Galactic sources of TeV gamma rays, and a test for neutrino emission from the {\it Fermi} Bubbles (large areas of diffuse gamma-ray emission observed above and below the Galactic Center)~\cite{fermi_bubbles}. 
We also performed an all-sky point-like source search and a test for emission from a catalog of known GeV (mostly extra-galactic) gamma-ray emitters. The results for each test~\cite{note:supplmat} are summarized in Table \ref{tab:results}.

\subsection*{Galactic plane neutrino searches}

We tested three models of Galactic diffuse neutrino emission, extrapolated from the observations in gamma rays (Figure~\ref{fig:gp_inserts}B).  These models are referred to as $\pi^0$, KRA$_\gamma^5$, and KRA$_\gamma^{50}$~\cite{kra} and are each
derived from the same underlying gamma-ray observations~\cite{fermi}.  
The model predictions depend on the distribution and emission spectrum of cosmic-ray sources in the Galaxy, the properties of cosmic-ray diffusion in the interstellar medium as well as the spatial distribution of target gas.  
Each neutrino emission model is converted to a spatial template, then convolved with the detector acceptance and the 
event's estimated angular uncertainty, to produce an event-specific spatial probability density function (shown for a typical event angular uncertainty of 7$^\circ$ in Figure~\ref{fig:gp_inserts}D). 

The $\pi^0$ model assumes that the MeV-to-GeV $\pi^0$ component, inferred from the gamma-ray emission, follows a power law in photon energy (E) of E$^{-2.7}$ and can be extrapolated to TeV energies with the same spatial emission profile.  The  KRA$_\gamma$ models 
include a variable spectrum in different spatial regions, use a harder (on average) neutrino spectrum than the $\pi^0$ model, and include a spectral cutoff at the highest energies~\cite{kra}. 
In this analysis, the KRA$_\gamma$ models are tested with a template that uses a constant, model-averaged spectrum over the sky, roughly corresponding to an E$^{-2.5}$ power law, with either a 5\,PeV or a 50\,PeV cosmic-ray energy cutoff for the KRA$_\gamma^5$ and KRA$_\gamma^{50}$ models respectively. The KRA$_\gamma$ models predict more concentrated neutrino emission from the Galactic Center region, whereas the $\pi^0$ model predicts events more evenly distributed along the Galactic plane.  The corresponding neutrino spectrum predicted by these models has a cutoff at about 10 times lower energies.

 Our Galactic template searches are performed with the same methods as previous Galactic diffuse emission searches \cite{gptracks, mesecascades }. Due to the uncertainties in the expected distribution of sources, their emission spectrum and cosmic-ray diffusion, we make no assumption about the absolute model normalization.  Instead, the analyses have an unconstrained parameter for the number of signal events ($n_s$) in the entire sky,  providing the flux normalization, while keeping the spectrum fixed to the model.  
 Results for each model are summarized in Table \ref{tab:results}.  We reject the background-only hypothesis with significance of 4.71$\sigma$, 4.37$\sigma$, and 3.96$\sigma$ for the $\pi^0$, KRA$_\gamma^5$ and KRA$_\gamma^{50}$ models respectively. Although these three hypotheses are correlated, we apply a conservative trial factor of three to the most significant of the three Galactic plane templates. The trial-corrected p-value results in a significance of 4.48$\sigma$.
 
\begin{figure}
    \centering
    \includegraphics[width=0.99\textwidth]{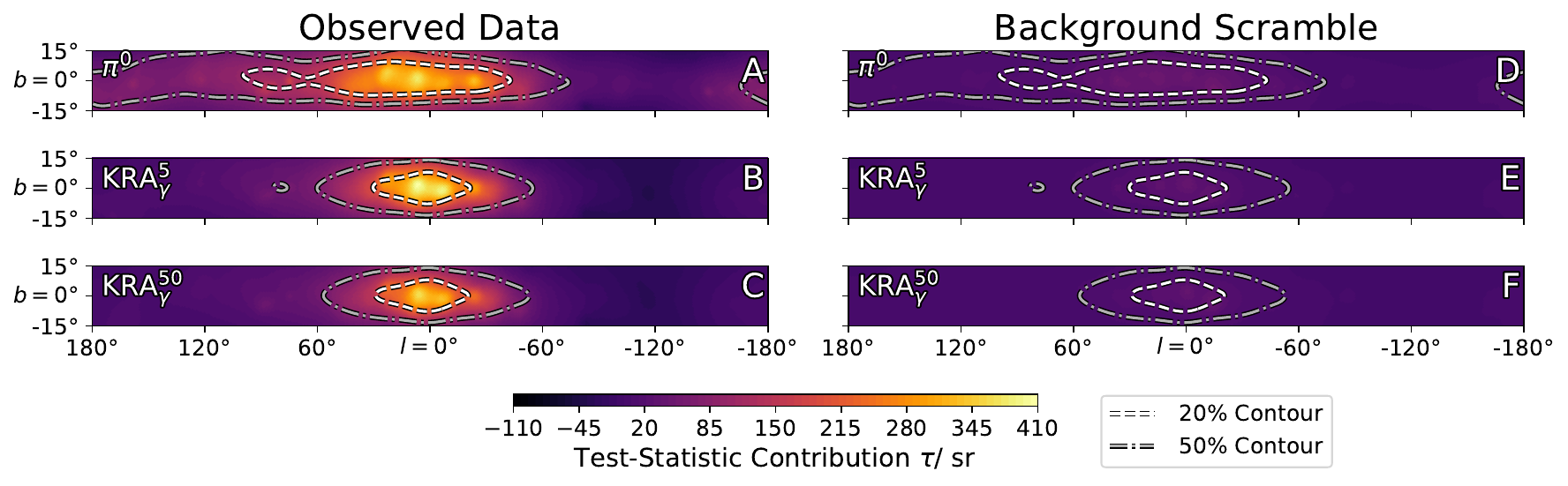}
    \caption{\textbf{Galactic plane test-statistic contributions.} The contribution to the test-statistic~$\tau$ is shown in galactic coordinates (longitude and latitude indicated by $l$ and $b$, respectively) for each of the three tested Galactic plane models. The overall test-statistic value was obtained by integration over the sky. The contribution for the observed data (A-C) is compared to the contribution for a single randomly selected mock experiment using scrambled data (D-F). 
    Contours enclose 20\% (white) and 50\% (gray) of the predicted model flux; for the $\pi^0$ model these are the same as in Figure~\ref{fig:gp_inserts}D-E. 
    The 50\% contours contain about
    $\SI{1.64}{sr}$, $\SI{0.70}{sr}$ and $\SI{0.65}{sr}$
    for the $\pi^0$, KRA$_\gamma^5$ and KRA$_\gamma^{50}$ models, respectively.
    }
    \label{fig:ts_contribution}
\end{figure}

The best-fitting fluxes are also listed in Table \ref{tab:results}. The flux normalization of the $\pi^0$ model is quoted at 100 TeV assuming a single power law; however, the KRA$_\gamma$ models have a more complex spectral prediction and are therefore quoted as multiples of the predicted model flux. These fluxes correspond to  best-fitting values of 748, 276, and 211 signal events ($n_s$) in the IceCube dataset for the $\pi^0$, KRA$_\gamma^5$ and KRA$_\gamma^{50}$ models, respectively. 
A visualization of the template results is provided in Figure~\ref{fig:ts_contribution}A-C, which shows the map of the per-steradian contribution to the results in the sky region near the Galactic Center for each of the Galactic plane models.  Similar maps for a randomly selected mock experiment are also shown for comparison (Figure~\ref{fig:ts_contribution}D-F).

\begin{figure}
    \centering
    \includegraphics[width=0.99\textwidth]{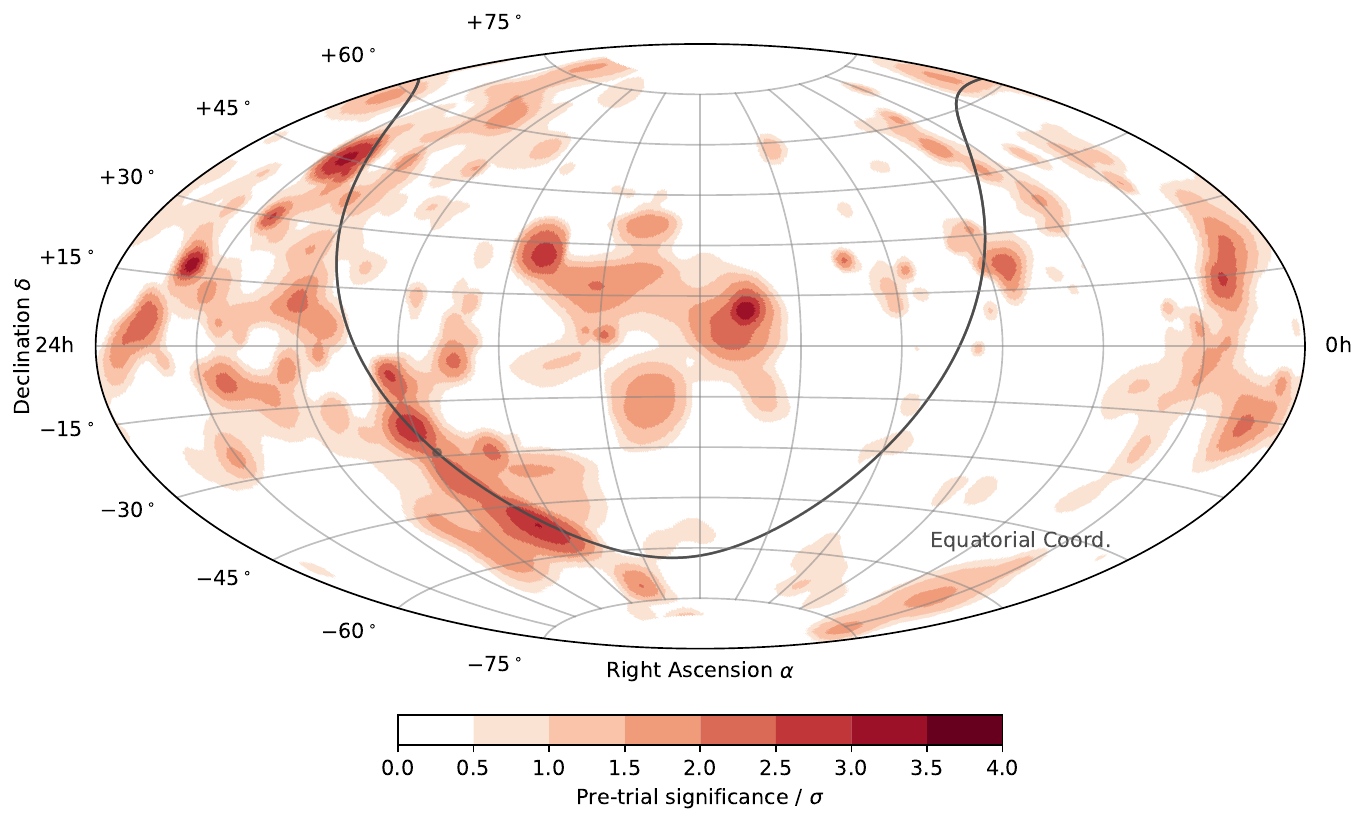}
    \caption{\textbf{All-sky point source search.} The best-fitting pre-trial significance for the all-sky search is shown as a function of direction in an Aitoff projection of the celestial sphere, in equatorial coordinates (J2000 equinox). The Galactic plane is indicated by a grey curve, and the Galactic Center as a dot. Although some locations appear to have significant emission, the trial factor for the number of points searched means these points are all individually statistically consistent with background fluctuations. }
    \label{fig:skymap_allsky}
\end{figure}

An all-sky point source search was also performed, in which the sky is divided into a grid of equal solid angle bins, spaced 0.45$^\circ$ apart, and each point is tested as a neutrino point source. 
The resulting significances are shown in Figure~\ref{fig:skymap_allsky}. Some locations have excess emission over the background expectations, including some in spatial coincidence with known gamma-ray emitters, such as the Crab Nebula, 3C 454.3, and the Cygnus X region. However, after accounting for trials factors, no single point in the map is statistically significant (Table~\ref{tab:results}).
This also implies that the emission that is present in the Galactic template analyses is not due to a single point source.

\subsection*{Searches using catalogs of Galactic sources}

The total gamma-ray signal from the Galactic plane includes a contribution from several strong gamma-ray point sources~\cite{fermi}.  
We therefore searched  for correlated neutrino emission from three distinct catalogs of Galactic sources, which previous work had classified as supernova remnants (SNR), pulsar wind nebulae (PWN), or other unidentified (UNID) Galactic sources based on observations in TeV gamma rays~\cite{TeVCat2-0, gamma-cat}. Under the assumption that multiple sources in each class emit neutrinos, stacking these sources in a single analysis provides higher sensitivity compared to individual source searches, because the neutrino fluxes add together while random background adds incoherently~\cite{IceCube:2010nca}.  The objects in each catalog were selected based on the observed gamma-ray emission above 100\,GeV and the detector sensitivity following methods described previously~\cite{10yrtracks}.  The 12 sources with the strongest expected neutrino flux are chosen from each category and  weighted under the hypothesis that they contribute equally to the flux~\cite{note:supplmat}. The total number of signal events and the spectral index are left as free parameters for each catalog search.  
The resulting p-value for each catalog search is shown in Table \ref{tab:results}. Each result rejects the background-only hypothesis at the 3$\sigma$ level or above. 
However, we can not interpret these neutrino event excesses as a detection, as the objects in these Galactic source catalogs overlap spatially with regions predicting the largest neutrino fluxes in the Galactic plane diffuse emission searches.

\section*{Implications of Galactic neutrinos}
\begin{figure}
    \centering
    \includegraphics[width=0.95\textwidth]{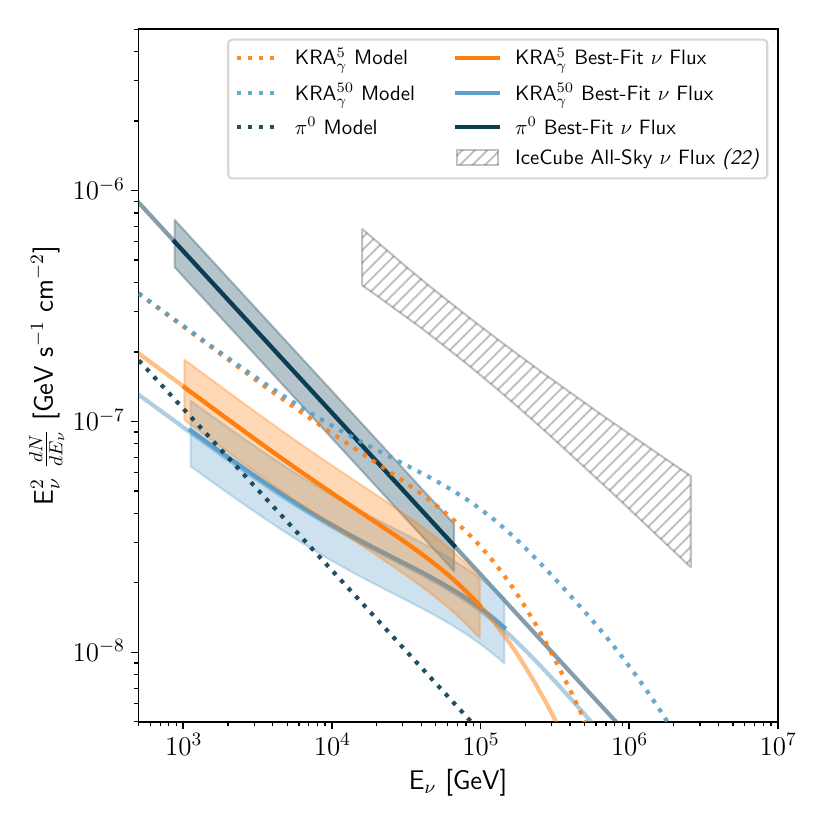}
    \caption{\textbf{Energy Spectra for each of the galactic plane models.} Energy-scaled, sky-integrated, per-flavor neutrino flux as a function of neutrino energy (E$_\nu$) for each of the Galactic plane models. Dotted lines are the predicted values for the $\pi^0$ (dark blue), KRA$_\gamma^5$ (orange) and KRA$_\gamma^{50}$ (light blue) models while solid lines are our best-fitting flux normalizations from the IceCube data. Shaded regions indicate the 1$\sigma$ uncertainties, extending over the energy range that contributes to 90$\%$ of the significance.  These results are based on the all-sky (4$\pi$ sr) template and are presented as an all-sky flux. For comparison, the grey hatching shows the flux of the IceCube all-sky neutrino flux~\cite{diffuse_cascades}, scaled to an all-sky flux by multiplying by 4$\pi$, with its 1$\sigma$ uncertainty.}
    \label{fig:gp_flux}
\end{figure}

The neutrino flux we observe from the Galactic plane could arise from several different emission mechanisms.
Figure~\ref{fig:gp_flux} shows the predicted energy spectra integrated over the entire sky for each of the Galactic plane models and their best-fitting flux normalization.  
Model-to-model flux comparisons depend on the regions of the sky considered. 
The KRA$_\gamma$ best-fitting flux normalizations are lower than predicted, which could indicate a spectral cutoff that is inconsistent with the 5\,PeV and 50\,PeV values assumed.
The simpler extrapolation of the $\pi^0$ model from GeV energies to 100\,TeV predicts a neutrino flux that is a factor of $\sim$5 below our best-fitting flux.   However, the $\pi^0$ model best-fitting flux appears consistent with recent observations of 100 TeV gamma rays by the Tibet Air Shower Array~\cite{TibetASgamma:2021tpz} (Figure~\ref{fig:gp_flux_comparison}).  The $\pi^0$ model mismatch could arise from propagation and spectral differences for cosmic rays in the Galactic Center region or from contributions from unresolved neutrino sources.

We use model injection tests to quantify the ambiguity between different source hypotheses. In these tests, the best-fitting neutrino signal from one source search is simulated, then the expected results in all other analyses are examined.
Injecting a signal from the $\pi^0$ model analysis, with a flux normalization equal to the best-fitting value from the observations, produces a median significance that is consistent with the best-fitting values for all other tested hypotheses within the expected statistical fluctuations. This includes the 3$\sigma$ excess observed in Galactic source catalog searches.
Individually injecting the best-fitting flux of any one of the tested Galactic source catalogs at the flux level observed does not recover the observed $\pi^0$ or KRA$_\gamma$ model results.  However, the angular resolution of the sample and the small number of equally weighted sources included in these catalogs does not constrain emissions from these broad source populations.  It is plausible that many independently contributing sources from the Galactic plane could show a comparable result to diffuse emission from interactions in the interstellar medium. In summary, these tests favor a neutrino signal from Galactic plane diffuse emission, but we do not have sufficient statistical power to differentiate between the tested emission models or identify embedded point sources.

The neutrinos observed from the Galactic plane  contribute to the all-sky astrophysical diffuse flux previously observed by IceCube~\cite{IceCube:2021uhz,diffuse_cascades,IceCube:2020wum}
(Figure~\ref{fig:gp_flux}).
The fluxes we infer for each of the Galactic template models contribute $\sim$6--13\% of the astrophysical flux at 30\,TeV.
These comparisons are complicated due to different spectral assumptions and tested energy ranges used in each analysis. Additionally, the observed Galactic flux is integrated over the entire sky, but local flux contributions along the central region of the Galactic Plane will be higher.

The observed excess of neutrinos from the Galactic plane provides strong evidence that the Milky Way is a source of high-energy neutrinos.  This evidence confirms our understanding of the interactions of cosmic rays within the Galaxy, as established by gamma-ray measurements, and complements IceCube's measurement of the diffuse extra-galactic flux to provide a more complete picture of the neutrino sky.

\begin{figure}
    \centering
    \includegraphics[width=.9\textwidth]{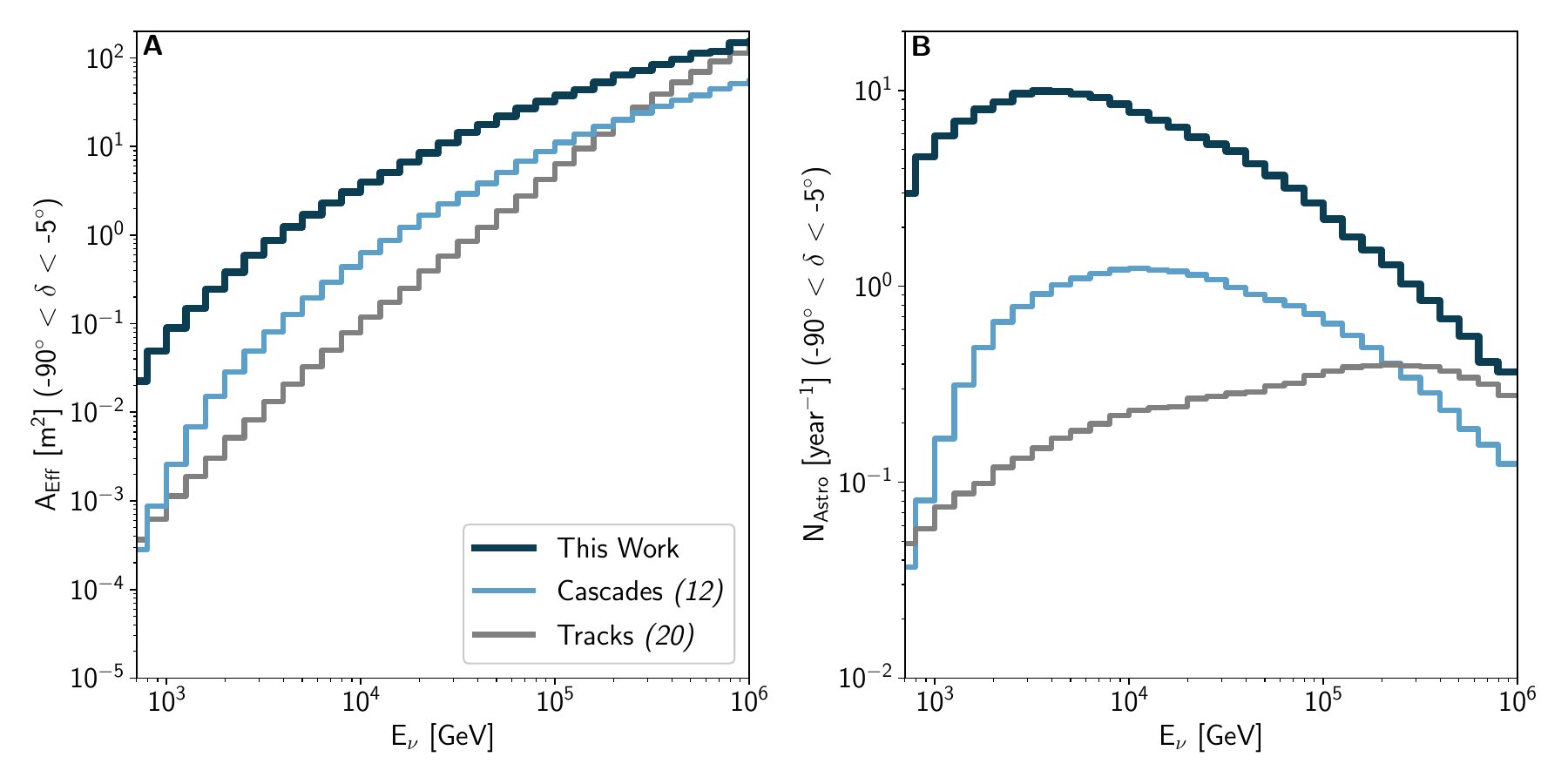}
    \caption{\textbf{Neutrino effective area and event selection comparison.} (A) The all-flavor Southern sky effective area (A$_{\rm Eff}$) of the IceCube dataset, averaged over solid angle in the declination ($\delta$) range between $-90^\circ$ and $-5^\circ$ as a function of E$_\nu$, the true neutrino energy.  Results are shown for the deep learning event selection used in this work, (dark blue), a previous cascade event selection~\cite{mesecascades} (light blue), and a previous track event selection~\cite{10yrtracks} (grey) applied to the IceCube data. (B) The number of expected signal events (N$_{Astro}$) in the Southern sky per energy bin per year for each event selection, assuming an isotropic astrophysical flux~\cite{diffuse_cascades}. Calculations are based on equal contributions of each neutrino flavor at Earth due to neutrino oscillations.}
    \label{fig:effa}
    
\end{figure}

\begin{table}[ht]
    \centering
    \caption{\textbf{Summarized results of the neutrino emission searches.}  The flux sensitivity and best-fitting flux normalization ($\Phi$) are given in units of model flux (MF) for KRA$_\gamma$ templates and as E$^2$ $\frac{dN}{dE}$ at 100\,TeV in units of 10$^{-12}$ TeV\,cm$^{-2}$\,s$^{-1}$ for the $\pi^0$ analyses ($\frac{dN}{dE}$ is the differential number of neutrinos per flavor, N, and neutrino energy, E). P-values and significance are calculated with respect to the background-only hypothesis. Pre-trial p-values for each individual result are shown for the three diffuse Galactic plane analyses and three stacking analyses, and post-trial p-values are shown for the other analyses.
    Due to the spatial overlap of the stacking catalogs with the diffuse Galactic plane templates, strong correlations between these searches are expected. The asterisk (*) indicates significance values that are consistent with the diffuse Galactic plane template search results. More detailed results for each search are provided in Tables~\ref{tab:allsky}-\ref{tab:source_list}.}
    \begin{tabular}{ccc c}
    \hline
         $\begin{array}{cc}
              \textbf{Diffuse Galactic} \\
              \textbf{plane analyses} 
         \end{array}$ & \textbf{Flux sensitivity $\Phi$} & \textbf{p-value}   & \textbf{Best-fitting flux $\Phi$} \\
         \hline
    \rule{0pt}{2ex} $\pi^0$ &  5.98 &  1.26$\times$10$^{-6}$  (4.71$\sigma$) &  21.8 $^{+5.3}_{-4.9}$  \\
    \rule{0pt}{2ex}KRA$_\gamma^{5}$ & 0.16$\times$MF &  6.13$\times$10$^{-6}$  (4.37$\sigma$) &   0.55$^{+0.18}_{-0.15}\times$MF \\
    \rule{0pt}{2ex}KRA$_\gamma^{50}$ & 0.11$\times$MF &  3.72$\times$10$^{-5}$ (3.96$\sigma$) &  0.37$^{+0.13}_{-0.11}\times$MF \\
    \hline

        $\begin{array}{cc}
              \textbf{Catalog stacking} \\
              \textbf{analyses} 
         \end{array}$ &  & \textbf{p-value}   &   \\
    \hline
    SNR &   & 5.90$\times$10$^{-4}$ (3.24$\sigma$)$^*$ &    \\
    PWN &  &  5.93$\times$10$^{-4}$ (3.24$\sigma$)$^*$ &   \\
    UNID &  &  3.39$\times$10$^{-4}$ (3.40$\sigma$)$^*$ &  \\
    \hline
   $
    $ \textbf{Other analyses} &  & \textbf{p-value}  &   \\
    \hline
    {\it Fermi} bubbles &   & 0.06   (1.52$\sigma$) &   \\ 
    Source list &   &   0.22  (0.77$\sigma$) &    \\
    Hotspot (North) &   &  0.28  (0.58$\sigma$) &      \\
    Hotspot (South) &   &  0.46  (0.10$\sigma$) &     \\
    \hline
    \end{tabular}

    \label{tab:results}
\end{table}


\FloatBarrier
\newpage
\pagebreak



\renewcommand\refname{References and Notes}
\bibliography{scibib}

\bibliographystyle{Science}

\subsection*{Acknowledgments}
The IceCube Collaboration gratefully acknowledges the contributions from D. Gaggero, D.
Grasso, A. Marinelli, A. Urbano, and M. Valli in providing the templates for the KRA-$\gamma$ models
and for fruitful discussions. \\

\noindent{\bf Funding:}
The authors gratefully acknowledge the support from the following agencies and institutions:
USA {\textendash} U.S. National Science Foundation-Office of Polar Programs,
U.S. National Science Foundation-Physics Division,
U.S. National Science Foundation-EPSCoR,
Wisconsin Alumni Research Foundation,
Center for High Throughput Computing (CHTC) at the University of Wisconsin{\textendash}Madison,
Open Science Grid (OSG),
Extreme Science and Engineering Discovery Environment (XSEDE),
Frontera computing project at the Texas Advanced Computing Center,
U.S. Department of Energy-National Energy Research Scientific Computing Center,
Particle astrophysics research computing center at the University of Maryland,
Institute for Cyber-Enabled Research at Michigan State University,
and Astroparticle physics computational facility at Marquette University;
Belgium {\textendash} Funds for Scientific Research (FRS-FNRS and FWO),
FWO Odysseus and Big Science programmes,
and Belgian Federal Science Policy Office (Belspo);
Germany {\textendash} Bundesministerium f{\"u}r Bildung und Forschung (BMBF),
Deutsche Forschungsgemeinschaft (DFG),
Helmholtz Alliance for Astroparticle Physics (HAP),
Initiative and Networking Fund of the Helmholtz Association,
Deutsches Elektronen Synchrotron (DESY),
and High Performance Computing cluster of the RWTH Aachen;
Sweden {\textendash} Swedish Research Council,
Swedish Polar Research Secretariat,
Swedish National Infrastructure for Computing (SNIC),
and Knut and Alice Wallenberg Foundation;
Australia {\textendash} Australian Research Council;
Canada {\textendash} Natural Sciences and Engineering Research Council of Canada,
Calcul Qu{\'e}bec, Compute Ontario, Canada Foundation for Innovation, WestGrid, and Compute Canada;
Denmark {\textendash} Villum Fonden and Carlsberg Foundation;
New Zealand {\textendash} Marsden Fund;
Japan {\textendash} Japan Society for Promotion of Science (JSPS)
and Institute for Global Prominent Research (IGPR) of Chiba University;
Korea {\textendash} National Research Foundation of Korea (NRF);
Switzerland {\textendash} Swiss National Science Foundation (SNSF);
United Kingdom {\textendash} Department of Physics, University of Oxford.\\

\noindent{\bf Author contributions:}
The IceCube collaboration acknowledges the significant contributions to this manuscript from Mirco H{\"u}nnefeld and Stephen Sclafani. The IceCube Collaboration designed, constructed and now operates the IceCube Neutrino Observatory. Data processing and calibration, Monte Carlo simulations of the detector and of theoretical models, and data analyses were performed by a large number of collaboration members, who also discussed and approved the scientific results presented here. The manuscript was reviewed by the entire collaboration before publication, and all authors approved the final version. 
Contributions are in the areas of (a) Conceptualization, Formal Analysis and
Methodology (b) Data Curation and Resources (c) Funding Acquisition, Project
Administration (d) Investigation (e) Software (f) Supervision (g) Validation (h)
Writing - original draft and visualization (i) Writing - review and editing.
M. Ackermann (b, c, d), J. Adams (d, e, g), M. Ahlers (c, d, i), M. Ahrens (d), J.M. Alameddine (d), A. A. Alves Jr. (e), N. M. Amin (d), K. Andeen (b, c, d), T. Anderson (d), G. Anton (c, d)
% ...


\noindent{\bf Competing interests:} There are no competing interests to declare.\\

\noindent{\bf Data and materials availability:} The data and software used in this paper are available from the IceCube data archive at:\\
{\url{https://icecube.wisc.edu/science/data/neutrino-emission-galactic-plane}}\\
and from Zenodo~\cite{mirco_huennefeld_2022_7412035}.\\

\noindent{\bf List of Supplementary Materials}\\
Materials and Methods \\
Supplementary Text \\
Figures S1 - S12\\
Tables S1 - S5\\
References (41-62)\\

\pagebreak
\pagebreak
\setcounter{page}{1}
\renewcommand{\thepage}{S\arabic{page}}

\begin{center}
\Large{
Supplementary materials for:\\
Observation of high-energy neutrinos from the Galactic Plane
}
\end{center}

\subsection*{IceCube Collaboration$^{\ast}$:}

R.~Abbasi$^{17}$,
M.~Ackermann$^{61}$,
J.~Adams$^{18}$,
J.~A.~Aguilar$^{12}$,
M.~Ahlers$^{22}$,
M.~Ahrens$^{51}$,
J.~M.~Alameddine$^{23}$,
A.~A.~Alves Jr.$^{31}$,
N.~M.~Amin$^{43}$,
K.~Andeen$^{41}$
% ...
\\
\\
% ...
$^{12}$ Universit{\'e} Libre de Bruxelles, Science Faculty CP230, B-1050 Brussels, Belgium \\
% ...
$^{17}$ Department of Physics, Loyola University Chicago, Chicago, IL 60660, USA \\
$^{18}$ Department of Physics and Astronomy, University of Canterbury, Private Bag 4800, Christchurch, New Zealand \\
% ...
$^{22}$ Niels Bohr Institute, University of Copenhagen, DK-2100 Copenhagen, Denmark \\
$^{23}$ Department of Physics, TU Dortmund University, D-44221 Dortmund, Germany \\
% ...
$^{31}$ Karlsruhe Institute of Technology, Institute for Astroparticle Physics, D-76021 Karlsruhe, Germany  \\
% ...
$^{41}$ Department of Physics, Marquette University, Milwaukee, WI, 53201, USA \\
% ...
$^{43}$ Bartol Research Institute and Department of Physics and Astronomy, University of Delaware, Newark, DE 19716, USA \\
% ...
$^{51}$ Oskar Klein Centre and Department of Physics, Stockholm University, SE-10691 Stockholm, Sweden \\
% ...
$^{61}$ Deutsches Elektronen-Synchrotron, D-15738 Zeuthen, Germany \\
% ...
$^\ast$E-mail: analysis@icecube.wisc.edu
\\
\\
% ...
\newpage

\renewcommand\thefigure{S\arabic{figure}}    
\setcounter{figure}{0}    
\renewcommand\thetable{S\arabic{table}}    
\setcounter{table}{0}

\section*{Materials and Methods}
\subsection*{Event Selection and Reconstruction}

\begin{figure}
    \centering
    \includegraphics[width=.74\textwidth]{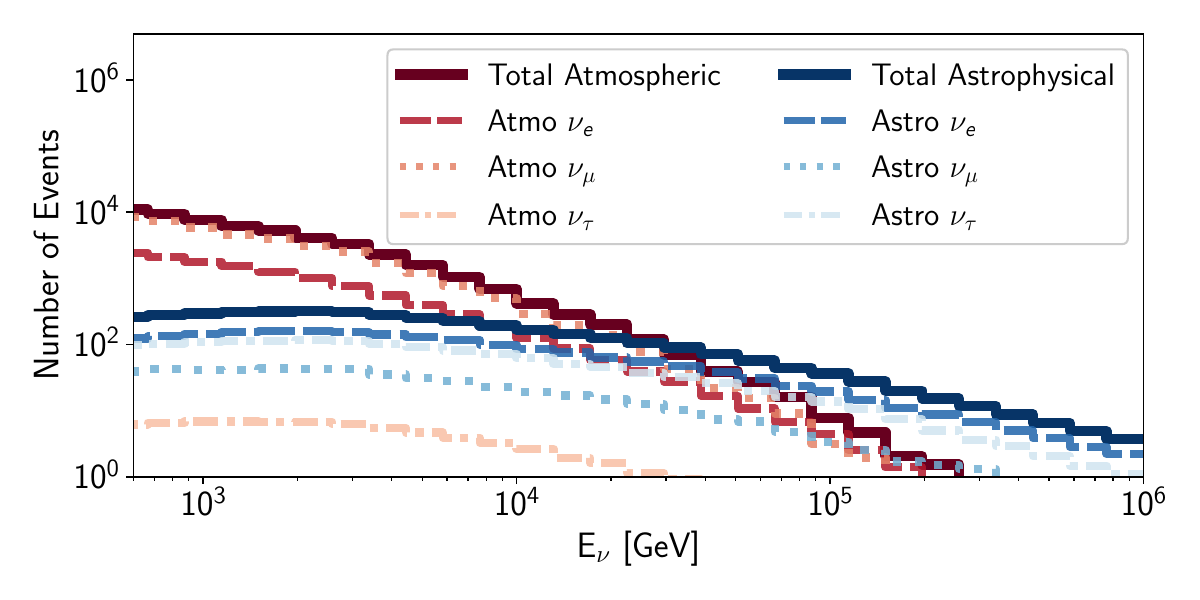}
    \caption{\textbf{Neutrino flavor event distribution.} Number of expected atmospheric neutrino (Atmo $\nu$) and astrophysical neutrino (Astro $\nu$) events for this work calculated based on simulation~\cite{7yrtracks}. Astrophysical neutrino contributions assume an isotropic astrophysical flux~\cite{diffuse_cascades} and an equal flavor ratio at Earth. The contributions of atmospheric neutrinos are obtained from the cascade-equation solver
MCEq~\cite{mceq}, the Hillas model (H3a) for cosmic rays~\cite{Gaisser:2011klf, Gaisser:2013bla} and the SIBYLL 2.3c
hadronic interaction model~\cite{sibyll}. Self-Veto \cite{Self-Veto-1, Self-Veto} passing fraction calculated with NuVeto \cite{nuveto}. Shown are the contribution from each flavor ($\nu$ + $\bar{\nu}$) as well as the total number of events as a function of neutrino energy (E$_\nu$).}
    \label{fig:Expected_events}
\end{figure}

\begin{figure}
    \centering
    \includegraphics[width=.8\textwidth]{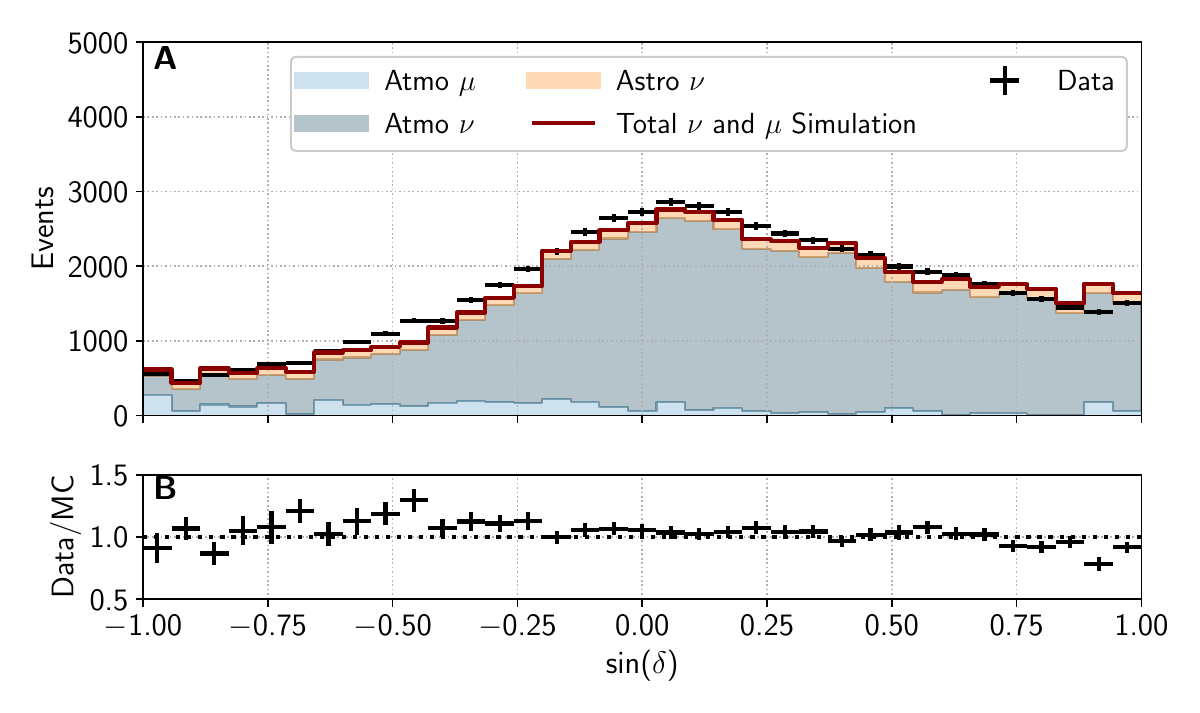}
     \includegraphics[width=.8\textwidth]{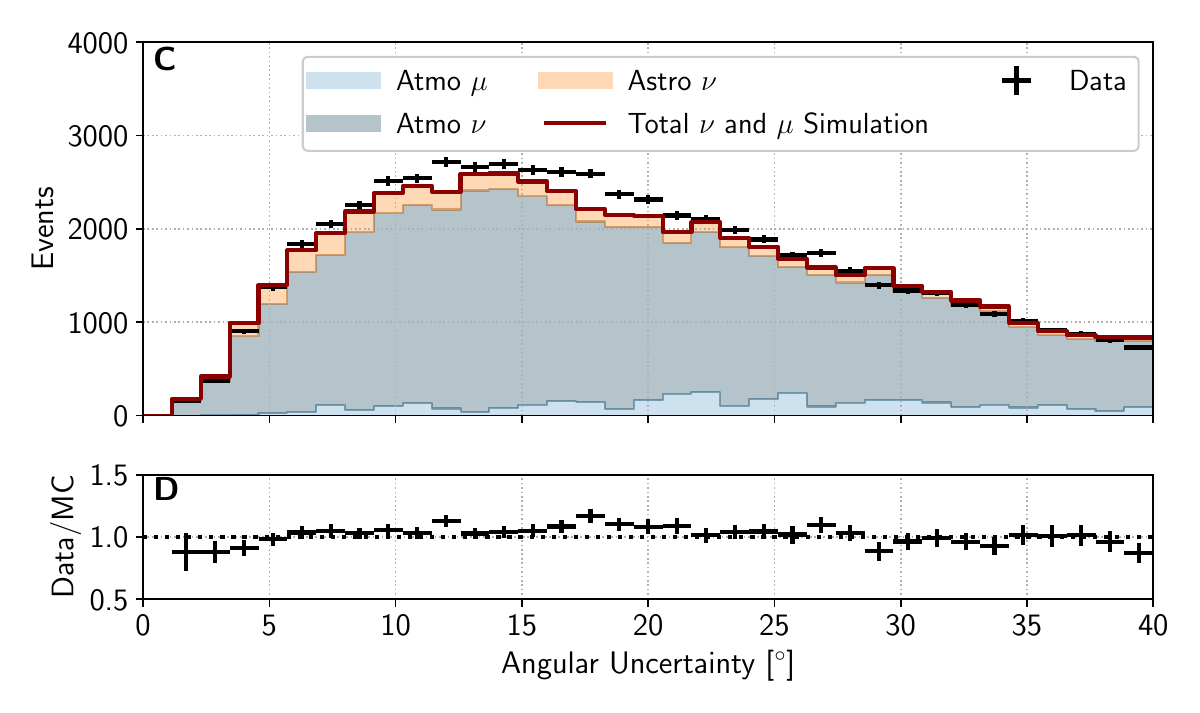}
    \caption{\textbf{Comparison between the data and Monte Carlo simulations.}
    Distributions of IceCube data and Monte Carlo simulation (MC)~\cite{7yrtracks} are compared for the reconstructed declination ($\delta$) (A) and the estimated, per-event angular uncertainty (C).
    Individual contributions to the simulation are illustrated as a stacked histogram with the atmospheric muon (Atmo $\mu$) atmospheric neutrino (Atmo $\nu$) and astrophysical neutrino (Astro $\nu$) components being added to reach the total of simulated signal and background events (red).  The astrophysical flux  (Astro $\nu$) is an isotropic power-law model fitted to IceCube data~\cite{diffuse_cascades}, with no additional Galactic component included.
    Panels B and D show the ratio of experimental data and simulation, including statistical 1$\sigma$ uncertainties from experimental data and limited Monte Carlo statistics.
    Systematic uncertainties are not included.
    }
    \label{fig:data_mc}
\end{figure}

The event selection utilizes a series of convolutional neural networks (CNNs) loosely based on the method presented previously~\cite{CNNPaper}. 
These CNNs include classification as well as regression tasks to obtain initial reconstructions for event properties such as direction, interaction vertex, and deposited energy. 
The event properties are later refined with a dedicated reconstruction method~\cite{Event-Generator}.
At early stages of the selection, fast and simple CNN architectures are applied to discard the majority of background events and thus reduce the data rate for subsequent selection steps. 
These CNNs require a per-event runtime of about $\SI{1}{ms}$ and they are able to reduce the atmospheric background by about 99.92\% more than the online cascade filter~\cite{diffuse_cascades, IceCubeDetector}, while retaining more than half of all signal events above 500 GeV. 
This reduction of background events by over three orders of magnitude allows for larger and more complex CNN architectures to be applied in subsequent steps. 
The final selection step is based on gradient boosted decision trees (BDTs)~\cite{XGBoost} that use high-level outputs of the CNNs as input.
At this stage, the atmospheric muon background is reduced by eight orders of magnitude and neutrinos  dominate the sample, thus enabling searches for neutrino sources. 
The per-flavor atmospheric and astrophysical neutrino contributions based on simulation~\cite{7yrtracks} are illustrated in Figure \ref{fig:Expected_events}.
Comparisons between experimental data distributions and simulation show reasonable agreement as illustrated in Figure~\ref{fig:data_mc}.
The analysis does not depend on an accurate modelling of the background due to the employed data-driven search method (see below).

\begin{figure}
    \centering
    \includegraphics[width=1.0\textwidth]{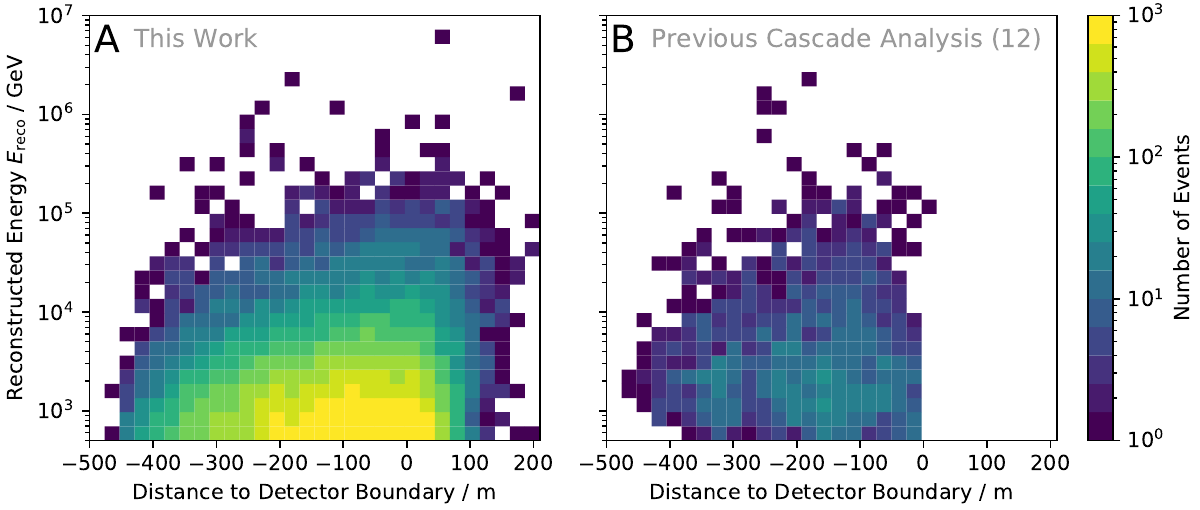}
    \caption{\textbf{Distribution of event vertex.} The distribution of experimental data events in reconstructed energy ($E_\mathrm{reco}$) and distance to the detector boundary are compared for this work (A) and the previous cascade selection~\cite{mesecascades} (B). 
    Negative distance values indicate that the interaction vertex is reconstructed inside the instrumented volume and thus these events are referred to as contained events.
    }
    \label{fig:containment}
\end{figure}

\begin{figure}
    \centering
    \includegraphics[width=0.45\textwidth]{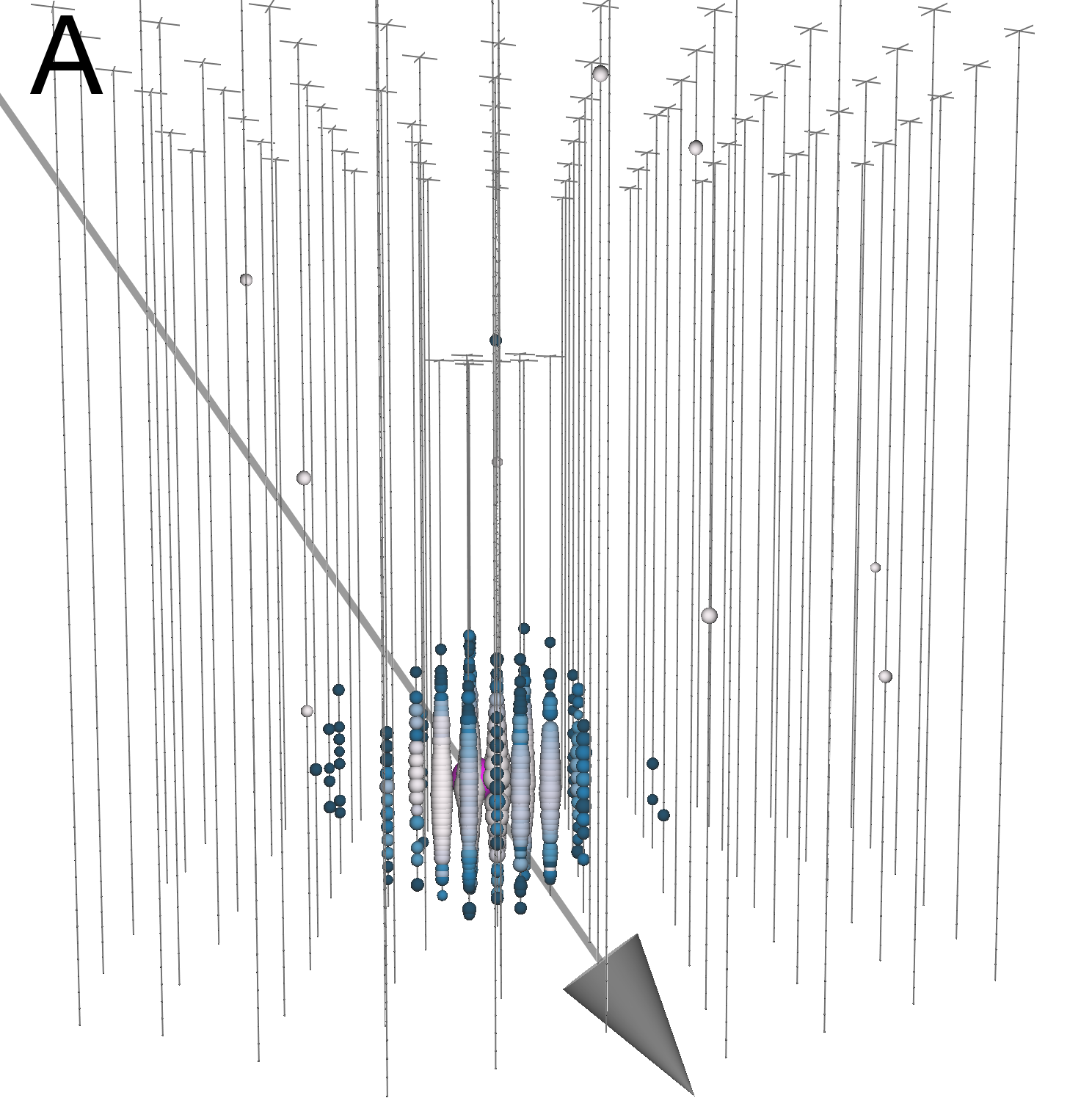}
    \includegraphics[width=0.45\textwidth]{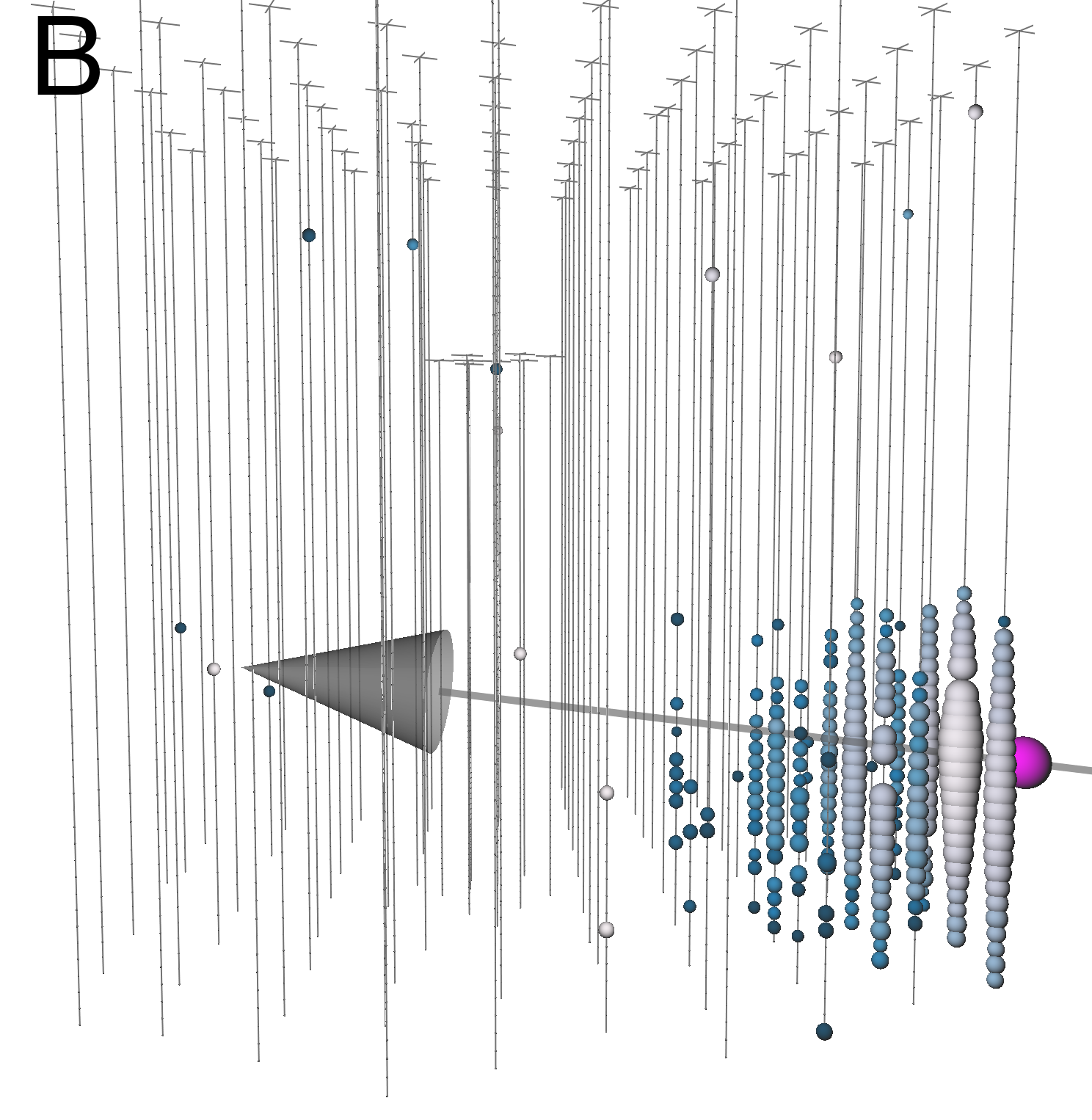}
    \caption{\textbf{Typical cascade event views.} Example event views of a ``contained'' (A) and an ``uncontained'' (B) cascade event. The colored blobs indicate DOMs that registered light, where the size of these blobs corresponds to the amount of observed light and the color indicates the time of the first registered light from early (white) to late (dark blue) times. Events such as these have time scales of a few $\mu$s. The magenta sphere and grey arrow show the reconstructed vertex and direction of the neutrino interaction. 
    }
    \label{fig:eventviews}
\end{figure}

\begin{figure}
    \centering
    \includegraphics[width=0.8\textwidth]{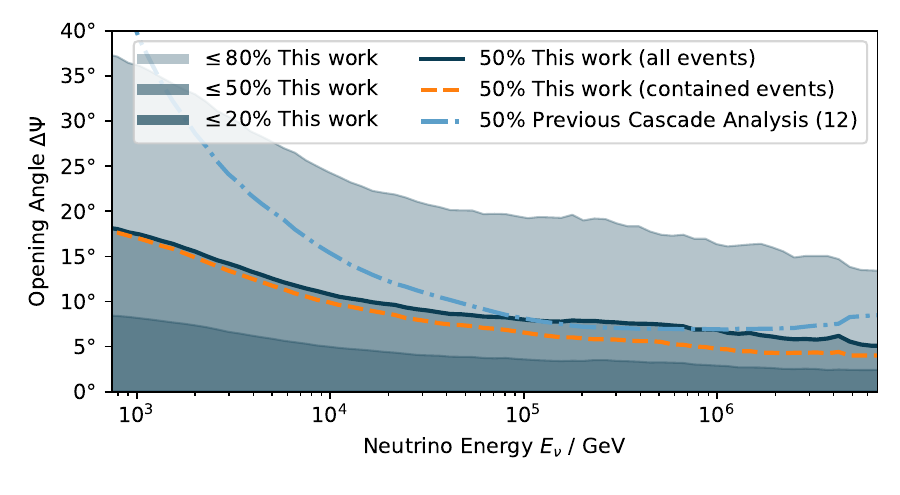}
    \caption{\textbf{Cascade event angular resolution.}  The angular resolution, defined as quantiles of the distribution of opening angles~($\Delta \Psi$) between true and reconstructed directions, as a function of neutrino energy ($E_\nu$) is shown for simulated events in this work (solid, black line and shaded regions) and the previous cascade selection~\cite{mesecascades}~(dashed-dotted). The dashed, orange curve shows the angular resolution of contained events. Systematic uncertainties are not included. }
    \label{fig:angular_resolution}
\end{figure}

\begin{figure}
    \centering
    \includegraphics[width=0.48\textwidth]{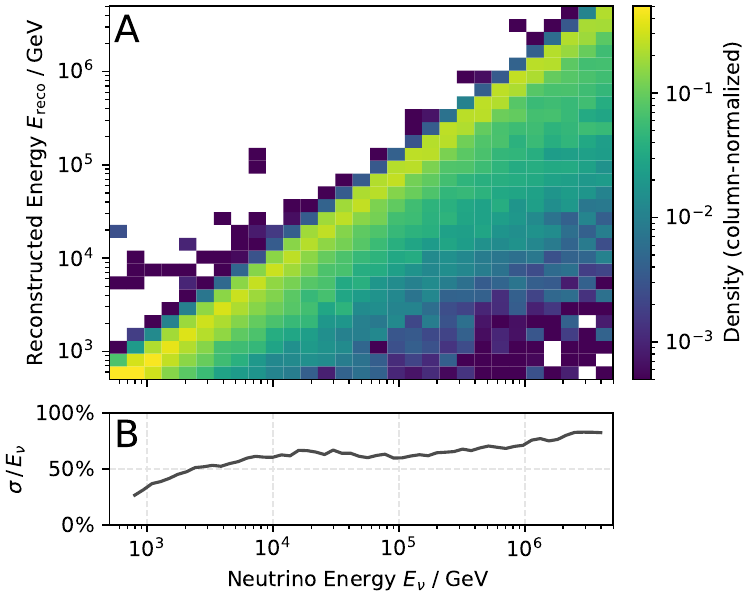}
    \hspace{12pt}
    \includegraphics[width=0.48\textwidth]{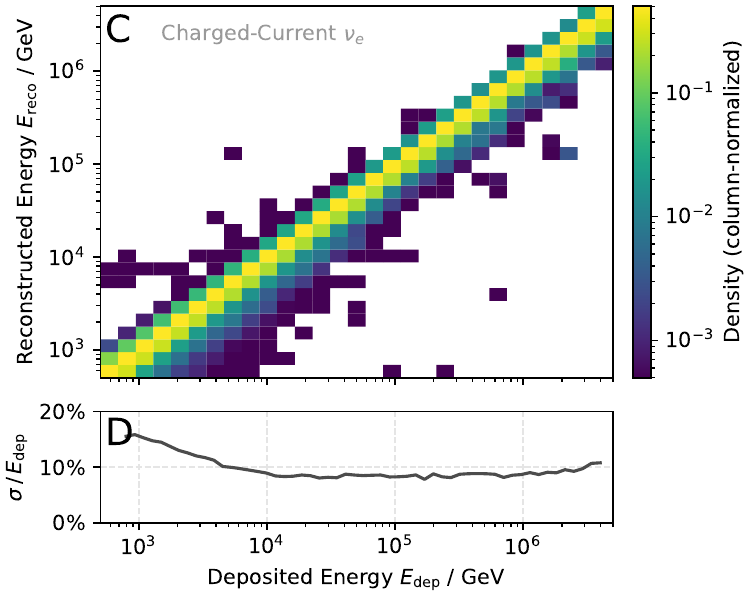}
 
    \caption{\textbf{Cascade event energy resolution.}  The energy resolution of simulated events is shown for the neutrino energy (A-B) and for the deposited energy of charged-current electron neutrinos (C-D). The reconstruction method aims to infer the true deposited energy ($E_\mathrm{dep}$) as shown in Panels C and D, which is a lower bound to the true neutrino energy ($E_\nu$)~\cite{EnergyReconstructionPaper}. Panels A and C illustrate the correlation between reconstructed~($E_\mathrm{reco}$) and true quantities. Panels B and D show the relative resolution, defined as the 68th percentile of the absolute values of the relative residuals $|(E_\mathrm{true} - E_\mathrm{reco}) / E_\mathrm{true}|$, where $E_\mathrm{true}$ is either $E_\nu$ (B) or $E_\mathrm{dep}$(D).}
    \label{fig:energy_resolution}
\end{figure}

The resulting cascade selection retains more than 20 times as many events as the previous IceCube  analysis~\cite{mesecascades}. 
This arises from both an overall improved efficiency and a lower energy threshold of about 500 GeV, compared to several TeV in earlier work, enabled by improved selection techniques.
A contribution to the increase
in efficiency relative to the previous selection~\cite{mesecascades} is the inclusion of events with interaction vertices near the boundary or outside of the instrumented volume as illustrated in Figure~\ref{fig:containment}.
These ``partially contained'' and ``uncontained'' events are more difficult to distinguish from background events and their reconstruction is challenging due to only observing a fraction of the deposited energy. Thus, these events were removed in most previous IceCube analyses.
The application of deep learning-based tools allows us to retain a large fraction of these and other challenging events, while ensuring reconstruction quality and a reduced level of background contamination.
An example for an ``uncontained'' and ``contained'' cascade event is shown in Figure~\ref{fig:eventviews}.
About 17.5\% of all events in the sample have a reconstructed interaction vertex outside of the instrumented volume and are thus considered as uncontained events.
An additional 2.5\% of events are located in a region with increased dust impurities in the ice.
Despite an increased fraction of more challenging cascade events, the energy-dependent angular resolution of the sample is improved over the previous selection as demonstrated in Figure~\ref{fig:angular_resolution}.
This is accomplished by the hybrid reconstruction method~\cite{Event-Generator}, which exploits more information than the CNN-based method~\cite{Huennefeld:2017pdh, CNNPaper} used in the previous cascade selection.
The energy resolution of this sample is illustrated in in Figure~\ref{fig:energy_resolution}.


\subsection*{Combining maximum-likelihood with deep learning}

The hybrid reconstruction method is a likelihood-based reconstruction algorithm that utilizes deep learning to approximate the underlying probability density function (PDF), i.e. the pulse arrival time distribution at each of the 5160 DOMs for any given light emitter-receiver configuration.
In previous  reconstruction methods~\cite{EnergyReconstructionPaper, IceCube:2021oqo}, this PDF was incorporated by dimensionality reductions and other approximations.
Our hybrid method uses neural networks to model these high-dimensional and complex dependencies.
It is constructed to exploit the  available physical symmetries and domain knowledge.
Details on how the neural network architecture was constructed, the training procedure, and the model's performance are provided elsewhere~\cite{Event-Generator}.

Due to the DOM's largely linear response to light intensity, any event hypothesis in IceCube can be deconstructed as a superposition of individual energy losses. The hybrid method makes use of this fact: only a single neural network (NN) needs to be trained to perform this elementary mapping. Complex event hypotheses can then be constructed from the superposition of multiple inference steps of this elementary NN. Since this analysis is focused on cascade events, most events are well modeled by a single energy deposition. However, fluctuations in energy depositions along the particle shower as well as incoming/outgoing particles and coincident events require a more complex event hypothesis. To this purpose, three event hypotheses are defined and reconstructed: a single energy deposition, two causally connected energy depositions, and two independent energy depositions. 
Of these three reconstructed hypotheses, the reconstruction result from the model hypothesis with the lowest estimated angular uncertainty is chosen for each event.  

The estimated per-event angular uncertainty for each of these reconstructed event hypotheses is obtained by three independent, fully-connected NNs.
These NNs are trained via a von Mises-Fisher likelihood~\cite{Fisher1953} to estimate the circularized angular uncertainty on each of these three reconstructions using high-level features such as the reconstructed event properties, the difference in  log-likelihood values between the hypotheses, and the second derivatives of the likelihood. This method provides better uncertainty estimation than by evaluating the Hessian of the likelihood at the best-fitting position. More sophisticated scans of the high-dimensional likelihood landscape were not performed due to computational constraints. 
This analysis utilizes circularized angular uncertainties, symmetrical in right ascension and declination. This assumption simplifies the analysis tools used for source searches, but it is not fully valid because cascades, in particular, can have asymmetric uncertainty contours, often elongated in right ascension due to the detector layout.

\begin{figure}
    \centering
    \includegraphics[width=0.48\textwidth]{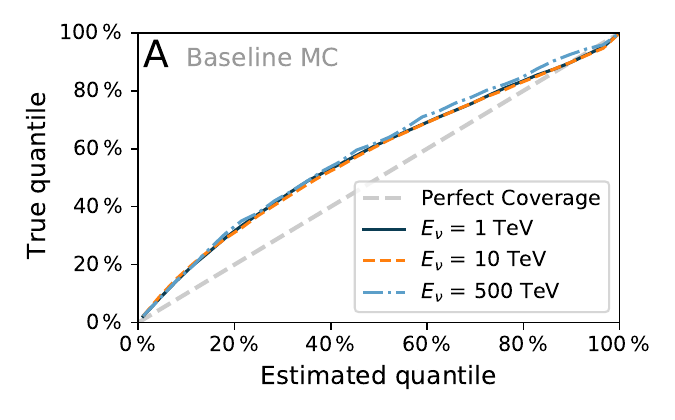}
    \hspace{12pt}
    \includegraphics[width=0.48\textwidth]{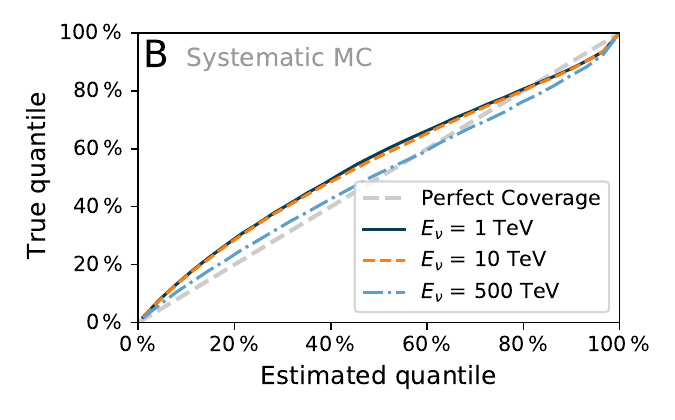}
    \caption{\textbf{Angular uncertainty estimator coverage.}
    The coverage of the angular uncertainty estimator is evaluated for different neutrino energies (colored lines) in the baseline MC simulation~(A) and in a simulation with varied systematic parameters~\cite{SnowStorm}~(B). 
    The angular uncertainty estimator provides an estimate for the fraction of events for which the true direction lies within a certain quantile (x-axis). This is compared to the true frequency at which the true direction falls into that quantile (y-axis). 
    Perfect coverage would fall along the diagonal indicated by the gray, dashed line.
    Curves that lie above this diagonal over-cover the true uncertainty, i.e. the estimated uncertainty is larger than the true uncertainty of the reconstruction method. 
    }
    \label{fig:coverage}
\end{figure}

Systematic uncertainties in the detector properties have an energy-dependent impact on the angular resolution, further degrading the resolution shown in Figure~\ref{fig:angular_resolution} by about 5\% at $\SI{1}{TeV}$ and about 30\% at $\SI{1}{PeV}$.
These systematic uncertainties encompass properties of the detector medium including absorption and scattering coefficients, ice anisotropy, and the acceptance parameterization of the refrozen ice column surrounding the DOMs, which has increased scattering and absorption properties due to enclosed air bubbles and dust impurities, as well as the DOM quantum efficiency. 
To account for these systematic effects, the estimated angular uncertainty is further corrected by an energy-dependent scaling factor that is determined from a systematic dataset that continuously samples from the estimated systematic uncertainties~\cite{SnowStorm}.
The coverage of the resulting uncertainty estimate is compared in Figure~\ref{fig:coverage} for the baseline MC simulation and the simulation with varied systematic parameters.
In the baseline simulation, the uncertainty estimate is over-covering by about 15\% due to the applied scaling factor. 
This over-coverage is reduced when introducing systematic uncertainties in the detector properties~(Figure~\ref{fig:coverage}B).
In both cases, the coverage does not strongly depend on the neutrino energy.
While an accurate uncertainty estimate is important, the computed p-values in this analysis are insensitive to this accuracy, and independent of any mis-modeling due to the employed data-driven search method (see below). 

\FloatBarrier

\subsection*{Spatial distribution of Galactic Models}
We use three models as templates for our diffuse neutrino search of the Galactic plane.
The Galactic plane models are based on spatial and spectral information derived from previous analyses of gamma-ray emission.  

The $\pi^0$ diffuse neutrino model is based on the $\pi^0$ gamma-ray component of the \textsc{GALPROP}~\cite{Strong:1998pw} output files available from the supplemental material for the  \texttt{$^S$S$^Z$4$^R$20$^T$150$^C$5} model~\cite{fermi}.
The $\pi^0$ gamma-ray component is then normalized to serve as a spatial probability template only, as in previous IceCube analyses \cite{mesecascades, gptracks}.

The KRA$_\gamma$ models~\cite{kra} are based on the same underlying gamma-ray observations as the $\pi^0$ model, but include a radially dependent diffusion coefficient for the cosmic-ray propagation, using the cosmic-ray transport software package \textsc{DRAGON}~\cite{DRAGON} and a diffuse gamma-ray simulation code \textsc{Gamma-Sky}~\cite{ GammaSky}. The files contain all-flavor neutrino flux simulations with a cosmic-ray cutoff of 5 PeV and 50 PeV~\cite{kra, gaggero_2022_7070823}. They were binned over the entire sky and energy range of the analysis and used as spatial and spectral templates. These files are identical to those used in previous IceCube analyses \cite{mesecascades, gptracks}.  Here, the model flux refers to the per-neutrino flavor, neutrino and anti-neutrino combined flux for each model.  

\begin{figure}
    \centering
    \includegraphics[width=0.99\textwidth]{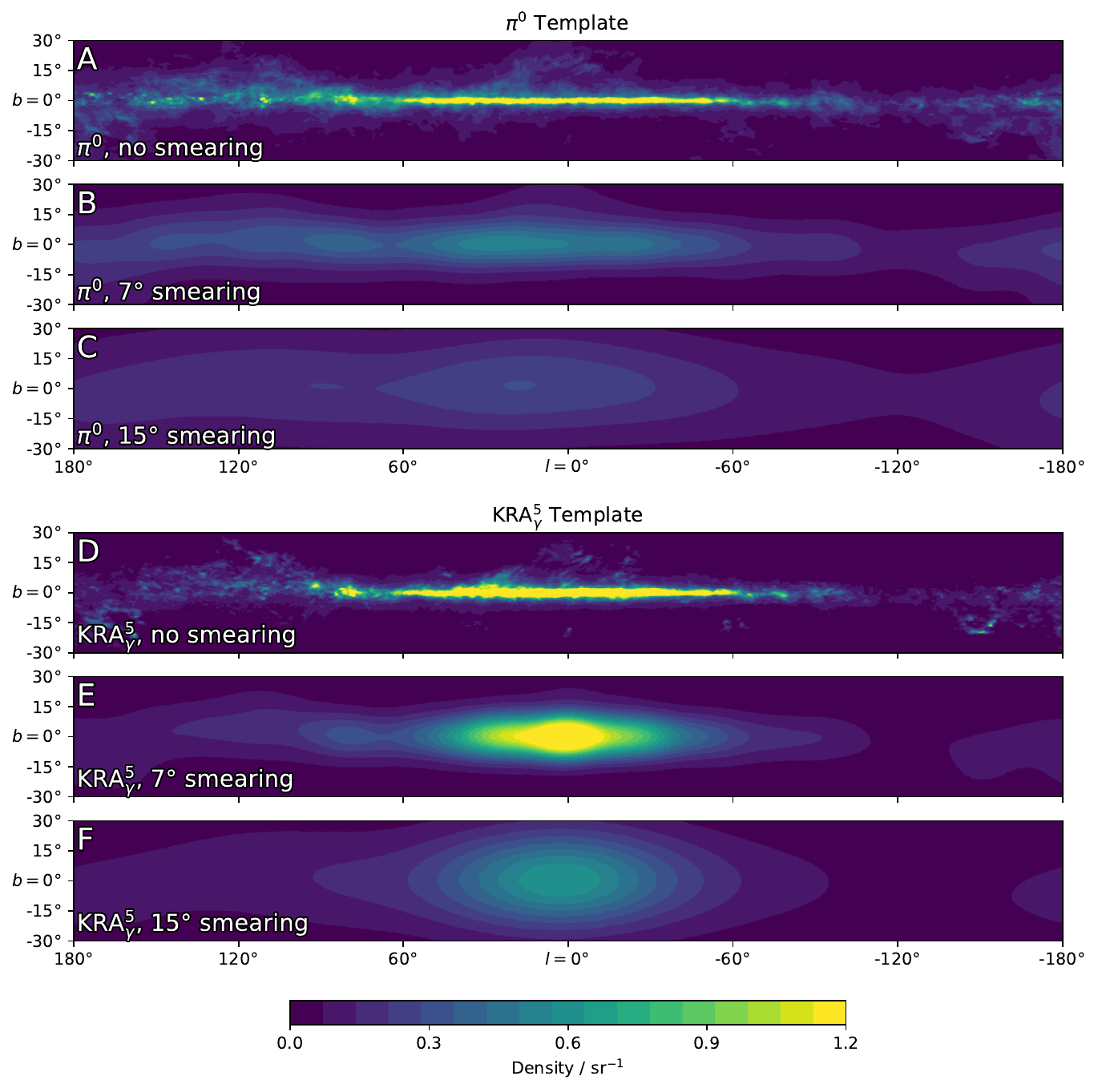}
    \caption{\textbf{Neutrino emission models used as templates in the Galactic plane search.} The spatial templates for the $\pi^0$ (A-C) and $\mathrm{KRA}_\gamma^5$ (D-F) models of diffuse Galactic neutrino emission are shown.
    Each panel shows the Galactic plane in a band of $\pm 30^\circ$ in latitude~($b$) and $\pm 180^\circ$ longitude~($l$) in Galactic coordinates.
    The models are first convolved with the IceCube detector acceptance (A, D) and then smeared with a Gaussian corresponding to the event uncertainty. Two example analysis templates are shown for a smearing of $7^\circ$ (B, E) and $15^\circ$ (C, F). The spatial distribution of the $\mathrm{KRA}_\gamma^{50}$ model is similar to the $\mathrm{KRA}_\gamma^5$ one shown here and it is available in the IceCube data archive.}
    \label{fig:templates}
\end{figure}

Figure~\ref{fig:templates} illustrates the Galactic template models after convolving with detector acceptance~(Figure~\ref{fig:templates}A and Figure~\ref{fig:templates}D) and additional Gaussian smearing of $7^\circ$~(Figure~\ref{fig:templates}B and Figure~\ref{fig:templates}E) and $15^\circ$~(Panels Figure~\ref{fig:templates}C and Figure~\ref{fig:templates}F).
At these resolutions, small scale structures visible in the spatial emission templates are washed out by the cascade directional resolution.
This analysis is therefore robust against mis-modeling of the spatial signal component of the Galactic emission models, by amounts up to a few degrees.
While this smearing is beneficial for the detection of unknown or poorly characterized extended sources, such as diffuse emission from the Galactic plane, it complicates further dissection of the signal observed in the emission region. 

\begin{figure} 
    \centering 
    \includegraphics[width=0.9\textwidth]{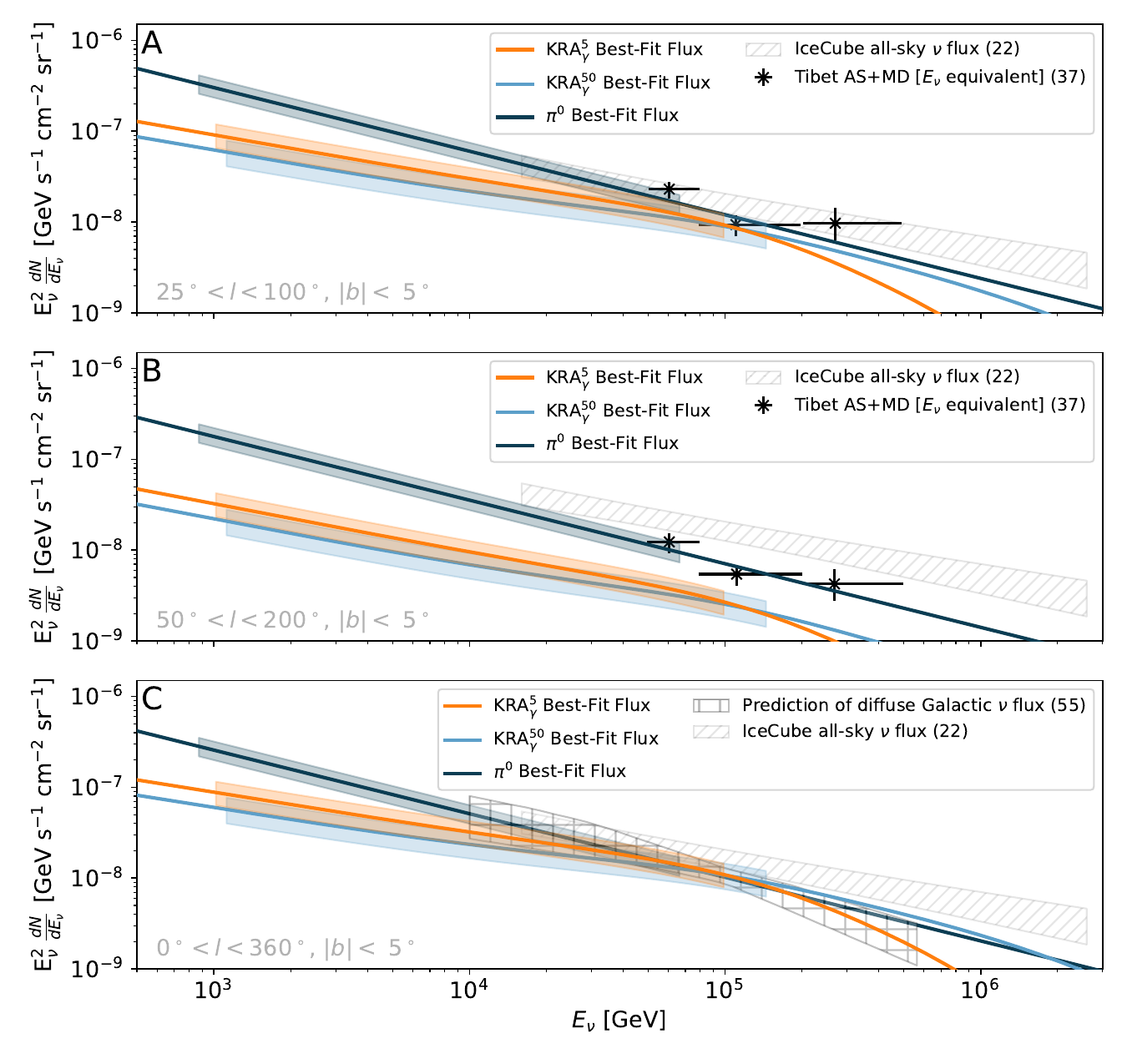}
    \caption{\textbf{Comparison between the best-fitting flux normalizations of the Galactic plane models.} 
    Same as Figure~\ref{fig:gp_flux}, but for flux averaged over three different regions of the sky.
    The average flux values are obtained by multiplying the total, sky-integrated neutrino flux from Table~\ref{tab:results} and Figure~\ref{fig:gp_flux} with the relative template contribution from each region, as indicated in the lower left of each panel.
    These fluxes are therefore not independent measurements in these parts of the sky, but an alternative presentation of the sky-integrated values. 
    Panels A-B include gamma-ray measurements from the Tibet Air Shower Array~\cite{TibetASgamma:2021tpz} (black asterisks), converted to a neutrino flux assuming a hadronuclear (pp) scenario~\cite{Murase:2013rfa, Ahlers:2013xia, Ahlers:2018fkn} neglecting gamma-ray attenuation.
    Panel C also shows a prediction for the diffuse Galactic neutrino flux~\cite{Fang:2021ylv} (checkered area), derived from gamma-ray measurements.
    }
    \label{fig:gp_flux_comparison} 
\end{figure}

The Galactic template searches fit each model to the integrated neutrino flux, taking the entire sky into consideration. 
The flux corresponding to a certain region of the sky is obtained by scaling the sky-integrated, best-fitting flux, illustrated in Figure~\ref{fig:gp_flux}, according to the relative contribution of that region based on the model template.
Figure~\ref{fig:gp_flux_comparison} shows such a comparison for three different Galactic plane regions together with gamma-ray measurements from the Tibet Air Shower Array~\cite{TibetASgamma:2021tpz} and neutrino predictions derived from gamma-ray measurements~\cite{Fang:2021ylv}.
The best-fitting neutrino fluxes do not constitute independent measurements for the specified sky regions, but rather a different presentation of the sky-integrated results.
This comparison indicates that the observed neutrino flux, when interpreted as a Galactic diffuse flux, is consistent with the flux level inferred by gamma-ray observatories in the TeV-PeV range.
However, contributions to the observed neutrino flux by unresolved sources cannot be constrained by our analysis. 

Due to spatial overlap of all tested neutrino emission models, combined with  the large angular uncertainty of the cascade events, correlations between the results of these model searches are expected.

\begin{figure}
    \centering
    \includegraphics[width=0.99\textwidth]{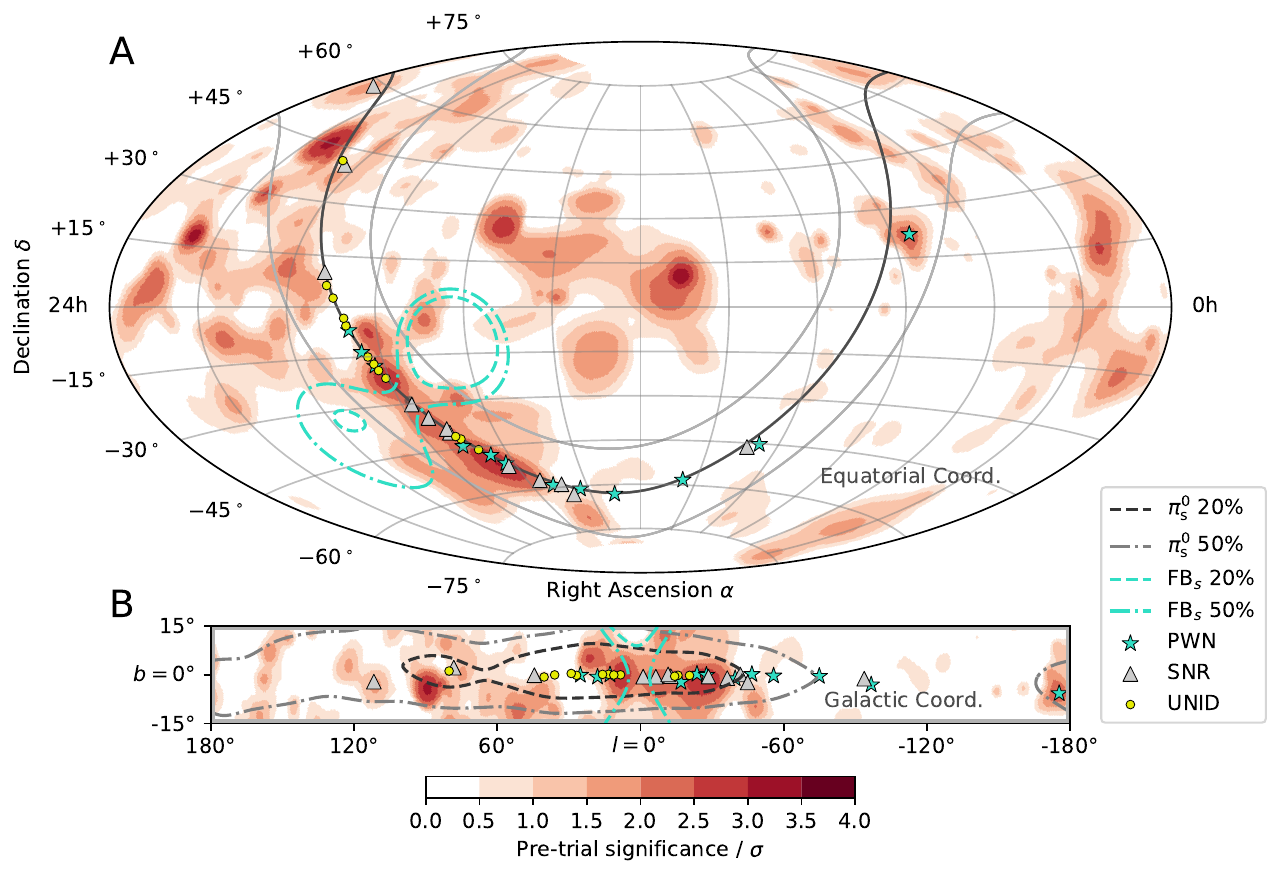}
    \caption{\textbf{All-sky search significance as a function of direction with tested sources.} 
    Same as in Figure~\ref{fig:skymap_allsky}, but with an additional $30^\circ$-cutout (indicated by grey lines) in galactic coordinates (longitude and latitude indicated by $l$ and $b$, respectively).
    Teal contours enclose 20\% and 50\% of the acceptance-corrected and smeared {\it Fermi} Bubbles template ($\mathrm{FB}_s$).
    Also shown are the sources of each of the three stacking catalogs, where the locations of sources are indicated by star, triangle, and circle symbols. The sources in the stacking catalogs follow the Galactic plane, indicated by a dark line. 
    The Galactic plane cutout (B) also shows the central 20\% and 50\% contours of the $\pi^0$ model ($\pi_s^0$) convolved with detector acceptance and smeared with a Gaussian corresponding to the uncertainty of a typical signal event (7$^\circ$),
    as shown in Figure~\ref{fig:gp_inserts}E.
    }
    \label{fig:skymap_significance}
\end{figure}

Figure~\ref{fig:skymap_significance} shows the pre-trial significance map of the all-sky search overlaid with the Galactic stacking catalog sources locations and {\it Fermi} Bubbles templates used in those source searches.
Individual event excesses are not statistically significant in excess of the background fluctuations as shown in Table~\ref{tab:results}.
However, there is an accumulation of warm spots along the Galactic plane and near the Galactic Center region, which is also densely populated with numerous sources in the catalog stacking searches.
The Galactic Center is also where the bulk of the Galactic diffuse emission is predicted by the $\pi^0$ and KRA$_\gamma$ templates (Figure~\ref{fig:templates}).

\FloatBarrier

\subsection*{Searches for Neutrino Emission}
\subsection*{Maximum Likelihood-Based Searches}
The search for neutrino sources, both point sources and extended, is based on the maximum-likelihood technique used in previous analyses of IceCube data~\cite{llhmethod}.  The signal hypothesis consists of neutrino emission from a particular source, following a power-law spectrum $E^{-\gamma}$ with spectral index ${\gamma}$.  
A likelihood function $\mathcal{L}$ is defined to represent the different hypotheses \cite{llhmethod},
\begin{equation}
     \mathcal{L}  (n_s, \gamma) = \prod_{i=1}^N \frac{n_s}{N} S_i(\gamma, \delta_i, \alpha_i, \sigma_i, E_i) + (1- \frac{n_s}{N})B_i (\delta_i, E_i),
    \label{eq:llh_trad}
\end{equation}
where $n_s$ is the number of signal events, $N$ is the total number of events, $S_i$ is the signal PDF of the $i$th event, which depends on declination ($\delta_i$), right ascension ($\alpha_i$), event angular uncertainty ($\sigma_i$), reconstructed energy ($E_i$), and spectral index ($\gamma$); $B_i$ is the per event background term which depends only on the declination and energy of each event.

Previous IceCube analyses, such as~\cite{10yrtracks}, assumed that the data in each declination/zenith range are background-dominated and the total event rate is a good approximation to the background. However, this assumption is not necessarily valid for our dataset. We therefore used an alternative method, described previously~\cite{gptracks}.  
This introduces a data-driven PDF, $\tilde{D}_i$, which does not correspond to pure background $B_i$, but includes the expected contribution from the signal PDF averaged over right ascension, $\tilde{S}_i$, in proportion to the strength of the signal that is being fitted:
\begin{equation}
\tilde{D}_i = (1-\frac{n_s}{N}) B_i + \frac{n_s}{N} \tilde{S}_i.
\end{equation}

The likelihood equation is then rearranged to arrive at the ``signal-subtracted" likelihood function:
\begin{equation}
 \mathcal{L} (n_s, \gamma) = \prod_{i=1}^N \frac{n_s}{N} S_i(\gamma, \delta_i, \alpha_i, \sigma_i, E_i) + \tilde{D_i} (\delta_i, E_i) - \frac{n_s}{N}\tilde{S_i} (\delta_i, E_i) 
    \label{eq:ssllh}
\end{equation}
The test statistic~$\tau$ is defined by the ratio of the maximized likelihood and the likelihood for the null hypothesis:
\begin{equation}
\tau = -2 \ln \frac{ \mathcal{L} (n_s = 0)}{ \mathcal{L}(\hat{n}_s,\gamma)}
    \label{eq:TS}
\end{equation}
where $\hat{n}_s$ is the best-fitting number of signal events.

This test statistic is calculated from the experimental data and converted to a p-value by comparing the experimental result to the test statistic distribution obtained from background mock-experiments generated from scrambled experimental data. To prevent bias, the reconstructed direction and energy of all events was kept blind and not considered during the development and characterization of these or subsequent analyses.  Due to IceCube's location at the South Pole and the rotation of Earth, backgrounds in this analysis (atmospheric events and an assumed-isotropic astrophysical neutrino background) are uniformly distributed in right ascension.
Similarly, the impact of any terrestrial sources of systematic uncertainty is also distributed evenly in right ascension.
The data-driven search method makes use of this, generating background mock-experiments by randomizing the right ascension values of the observed experimental data. 
This randomization removes signal from sources by evening out anisotropic structure on the sky (in given declination bands), while maintaining the influence of systematic uncertainties and the statistical properties of the background events in the experimental data.  This data-driven procedure was chosen to reduce the risk of false discovery of anisotropic structures on the sky. 

To account for multiple hypotheses tested in one analysis, such as the three similar Galactic plane emission templates, we convert the most significant of the p-values into a post-trial p-value by using a trial-factor.  This trial factor is equal to the number of tests for uncorrelated hypotheses, but can be lessened by taking correlations into account.
 
\subsection*{Signal Hypotheses}

We separate signal hypotheses into two components, $S_\mathrm{Space}$ and $S_\mathrm{Energy}$. These are defined per event and rely on the observables of reconstructed direction, energy, and angular uncertainty.  $S_\mathrm{Energy}$ exploits the different energy spectrum for astrophysical neutrinos relative to that of the background of atmospheric neutrinos and muons.  In general, higher energy events provide more support for the signal hypothesis, because the expected Galactic-plane neutrino flux has a spectrum that falls less steeply with energy than that of the atmospheric backgrounds.

$S_\mathrm{Space}$ is defined based on the expected signal hypothesis.  For a point source, $S_\mathrm{Space}$ is given for each event by a von Mises-Fisher distribution based on the event’s reconstructed direction ($x_i$), its estimated angular uncertainty ($\sigma_i$) and the source location ($x_s$).

For template searches, full sky templates are binned in equal-solid-angle bins, convolved with the detector acceptance, and smeared with a Gaussian with width equal to each event's angular uncertainty.  Each event therefore has a particular spatial weight based on its position, energy, direction, estimated angular uncertainty, and the signal hypothesis under investigation.  
These combined spatial and energy weights are multiplied and the likelihood is maximized for the analysis parameters described above.

\subsubsection*{Effective Area}

The effective area is defined as the declination and energy dependent area of a 100\% efficient detector and is calculated using simulations.  We calculate the effective areas as the sum of all neutrino flavors and types assuming equal flavor ratios. Effective areas in Figure~\ref{fig:effa} are presented as integrated over declination ranges and then divided over the solid angle to produce an averaged effective area for that region.
For a given flux $\Phi$, which is equal to both the neutrino and anti-neutrino fluxes, $\Phi$ = $\Phi_\nu$ =  $\Phi_{\bar{\nu}}$, the rate of observed neutrinos in time, $\frac{dN}{dt}$, is given as:

\begin{equation}
    \frac{dN}{dt} = \int d\Omega \int_{0}^\infty  A^{\nu+\bar{\nu}}_\mathrm{eff}(E_\nu, \delta) \times \Phi (E_\nu)\,dE_\nu ,
\end{equation}
where $A_\mathrm{eff}$ is the energy and declination dependent effective area and $\Omega$ denotes the solid angle.

\subsection*{Systematic Uncertainties and their Impact}

Here we consider the impact of systematic errors on our results.  The data-driven search method (see above) returns p-values that are derived from scrambled experimental data, thereby reducing the impact of un-modelled systematic uncertainties.  The performance of the event selection and reconstruction methods has been validated using simulation samples~\cite{7yrtracks}. The projected analysis sensitivities and effective areas, used to convert the fitted number of events~$n_s$ to a flux normalization, are based on simulations of the detector~\cite{7yrtracks}.   Therefore the estimated flux normalization is affected by systematic uncertainties, while the p-value remains robust.

Systematic uncertainties pertaining to ice-modeling (scattering, absorption, anisotropy, and properties of refrozen hole-ice column) as well as detector parameters such as the DOM quantum efficiency are accounted for in multiple places.
The CNNs in the event selection pipeline are trained on a variety of different simulation datasets with different systematic properties~\cite{SnowStorm}. The impact of these systematics on the CNNs has been quantified by previous work~\cite{CNNPaper}.
The neural network model utilized in the hybrid method is also trained on simulations that sample different sets of systematic parameters for each subset of events~\cite{SnowStorm}. For the loss function, used to evaluate prediction errors during training of the NN, the model utilizes a likelihood that incorporates the over-dispersion resulting from marginalization of systematic parameters~\cite{Event-Generator}.
The event reconstruction (see above) increases the per-event angular uncertainty estimates to account for these sources of systematic uncertainties.
The confidence interval construction of the best-fitting flux normalization of the Galactic plane models, reported in Table~\ref{tab:results} and Figure~\ref{fig:gp_flux}, also account for these systematic uncertainties. The confidence intervals are constructed by inversion of the likelihood ratio test~\cite{Feldman:1997qc} while also sampling a realisation of systematic parameters for every trial. As such, the confidence intervals not only include statistical uncertainties, but also known sources of systematic uncertainties. 

We find this set of systematic uncertainties impact the sensitivity by up to 20\% and the effective area by about 10\%. 
The largest impact is on the angular resolution of high-energy events, which are most affected by systematic uncertainties in the modeling of the glacial ice. Each of these effects are -- by construction -- accounted for in the p-value calculation.
Estimates of the best-fitting flux normalizations are susceptible to systematic uncertainties, but the identification of neutrino emission from the Galactic plane is robust due to the data-driven search method.

\newpage
\section*{Supplemental Text}
\subsection*{Additional Neutrino Source Searches}
\subsubsection*{All-Sky Scan}
A test for point source neutrino emission is performed over the full sky, searching for an excess of neutrino events over the background.  The method matches previous searches \cite{10yrtracks,mesecascades} where all directions in the sky are evaluated as a potential point source, by fitting the number of signal events $n_s$ and the power-law spectral index $\gamma$.  This test is evaluated on a grid of points of equal solid angle bins spaced 0.45$^\circ$ apart, looking for the brightest spot in each hemisphere.  Although individual points are highly correlated with their neighbors, this test still entails a large ($\sim$500) trial factor due to searching the whole sky.  The final results are reported in Table \ref{tab:allsky} as the trial-corrected p-value (p-value$_{\rm post}$) for the hottest point in the Northern and Southern Hemisphere. Both results are consistent with the background-only hypothesis.
\begin{table}
    \centering
    \caption{\textbf{All-sky search most significant locations.}  Results of the search for the most significant point in the Northern and Southern hemisphere.  Best-fitting number of signal events (n$_s$), spectrum ($\gamma$) pre-trial p-value (p-value$_{pre}$), as well as the trial corrected p-value (p-value$_{post}$) are presented. Other points in the sky are almost as significant.}
        \vspace{0.5em}
    \begin{tabular}{c|c|c|c|c|c|c}
         Analysis & $\alpha [^\circ]$ & $\delta [^\circ]$ & n$_s$ & $\gamma$ & p-value$_{\textsf{pre}}$ & p-value$_{\textsf{post}}$   \\
         \hline
         Hotspot (North) & 337.9 & 17.6 & 213.7 & 3.6 & 3.92$\times10^{-4}$ & 0.28 \\
         Hotspot (South) & 248.1 & -50.9 & 90.2 & 2.9 & 1.31$\times10^{-3}$ & 0.46 \\
         \hline
    \end{tabular}

    \label{tab:allsky}
\end{table}

\begin{figure} 
    \centering 
    \includegraphics[width=0.99\textwidth]{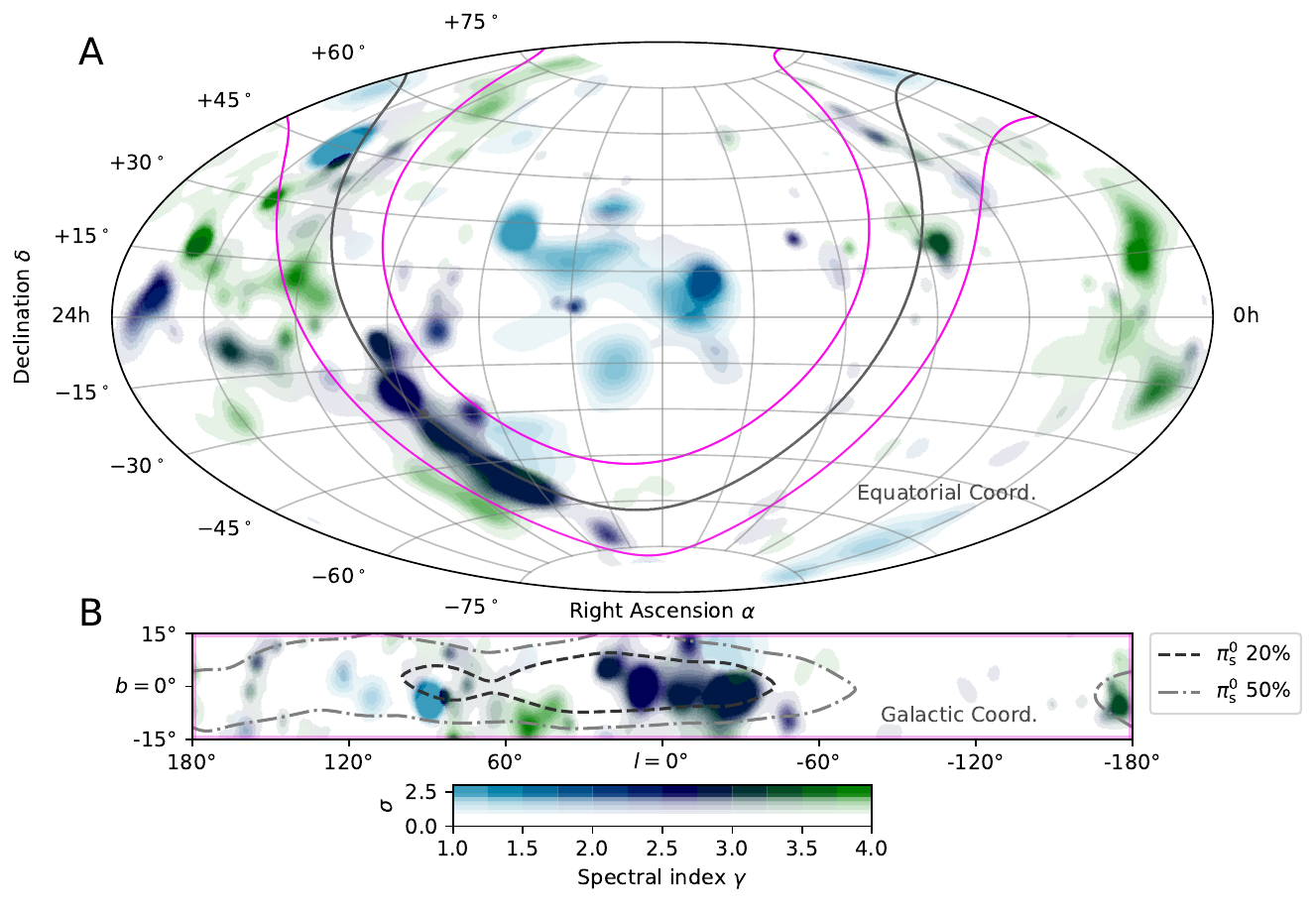}
    \caption{\textbf{All-sky search significance and spectral index as a function of direction.} 
    The best-fitting spectral index, weighted by pre-trial significance, is shown as a function of direction, in equatorial coordinates (J2000 equinox) and Aitoff projection, for the all-sky search. The pixel opacity is scaled by the pre-trial significance so more opaque locations are more significant. All excesses of neutrinos are consistent with background fluctuations, given the large trials factor. The Galactic plane is indicated by a grey curve with a magenta band, and the region between $\pm$15$^\circ$ in galactic latitude is highlighted in Panel B. Contours enclose 20\% and 50\% of the $\pi^0$ model convolved with detector acceptance and smeared with a Gaussian corresponding to the uncertainty of a typical signal event~(7$^\circ$).
    } 
    \label{fig:skymap_gamma_combined} 
\end{figure}

A skymap illustrating the best-fitting spectral index and significance at each location in the sky is provided in Figure~\ref{fig:skymap_gamma_combined}.
The all-sky map shows some warm spots in coincidence with nearby known gamma-ray sources, such as the Crab Nebula. 
However, for the point-source tests that were defined a priori (all-sky scan and source list search), no individual point-like source is statistically significant after accounting for the trial factor for the corresponding analysis.
The ability to spatially resolve a neutrino source  in the all-sky scan depends on the expected number of signal events and their energy distribution.
Due to the typical $5^{\circ}$ to $20^{\circ}$ angular uncertainty of individual cascade events, the combination of many such events is required to improve the ability to detect a source in the sky.
Assuming the source is a point source with parameter values as measured, we expect to resolve the source to a few degrees.

\subsubsection*{Fermi Bubbles and Source Stacking}

Neutrino emission from the {\it Fermi} Bubbles region around the Galactic Center was searched for.  The basic template used matches that in previous searches of IceCube data~\cite{mesecascades}, and is constructed by creating two circular lobes of radius 25$^\circ$ tangent to each other at the Galactic center.  Each point in those lobes is assumed equally likely to emit neutrinos following an $E^{-2}$ spectrum, with different spectral cutoffs being tested. The full results are shown in Table \ref{tab:fermibubble_results}, and are all consistent with background.  Multiple tests were performed and a trial corrected p-value of $\num{6.48e-2}$ (1.52$\sigma$) is obtained that accounts for the correlations between the different cutoffs, and this value is reported in Table \ref{tab:results}.

\begin{table}
    \centering
    \caption{\textbf{Fermi Bubble neutrino searches.}  Pre-trial significance, sensitivity, best-fitting number of signal events (n$_s$), and 90\% Upper Limits (UL) for {\it Fermi} Bubbles model with various exponential cutoffs. 
    The per-flavor neutrino flux sensitivity and upper limits are given as E$^2$ $\frac{dN}{dE}$ at 1\,TeV in units of 10$^{-12}$ TeV\,cm$^{-2}$\,s$^{-1}$.}
    \vspace{0.5em}
    \begin{tabular}{c|c|c|c|c|c}
    \hline
         Cutoff & Sensitivity $\Phi$ & n$_s$ & p-value & Significance ($\sigma$) & UL $\Phi$ \\
         \hline
         50\,TeV & 20.87  & 96.0 & 0.03 & 1.88 & \textless52.51 \\
         100\,TeV & 14.68 & 70.9 & 0.05 & 1.65 & \textless34.06 \\
         500\,TeV & 7.74 & 34.4 & 0.12 & 1.17 & \textless14.99 \\
         No Cutoff & 3.15 & 23.7 &  0.14 & 1.06 & \textless 5.91 \\ 
         \hline
    \end{tabular}

    \label{tab:fermibubble_results}
\end{table}

The full results for stacking catalogs are shown in Table \ref{tab:stacking_results}.  As described in the main text, the analysis searches for total emission from a catalog of point-like sources. Each source in a particular catalog is weighted to equally contribute to the flux before any detector effects are considered.  The analysis determines the best-fitting total number of signal events and therefore total flux and a single catalog power law spectral index ($\gamma$).  
\begin{table}
    \centering
    \caption{\textbf{Source catalog search summary results.}  Pre-trial significance, sensitivity, best-fitting spectrum ($\gamma$), total number of signal events (n$_s$), flux, and 90\% Upper Limits (UL) for the Galactic stacking catalog analyses.  Upper limits are with respect to a source emitting following an E$^{-2}$ spectrum. 
    The per-flavor neutrino flux sensitivity, best-fit, and upper limits are given as E$^2$ $\frac{dN}{dE}$ at 100\,TeV in units of 10$^{-12}$ TeV\,cm$^{-2}$\,s$^{-1}$ for the entire catalog of sources.}
    \vspace{0.5em}
    \begin{tabular}{c|c|c|c|c|c|c|c}
    \hline
         Catalog & Sensitivity $\Phi$ & n$_s$ &  $\gamma$ & p-value & Signficance ($\sigma$) & Flux $\Phi$ & UL $\Phi$ \\
         \hline
         SNR & 2.24 & 218.6 &  2.75 & 5.90$\times$10$^{-4}$ & 3.24 & 6.22 & \textless9.01 \\
         PWN & 2.25 & 279.6 & 3.00 & 5.93$\times$10$^{-4}$ & 3.24 & 3.80 & \textless9.50 \\
         UNID & 1.89 & 238.4 & 2.85 & 3.39$\times$10$^{-4}$ & 3.40 & 5.03 & \textless7.76 \\
         \hline
    \end{tabular}

    \label{tab:stacking_results}
\end{table}

\begin{table}
    \centering
       \caption{\textbf{Galactic source catalog.}  The sources and locations~\cite{gamma-cat, TeVCat2-0} of galactic sources used in stacking catalogs.}
    \begin{tabular}{c|c|c|c}
    \hline
     Catalog & Source & $\alpha$ [$^\circ$] & $\delta$ [$^\circ$] \\
     \hline
SNR & Vela Junior  & 133.0 & -46.33  \\
SNR & RX J1713.7-3946 & 258.36 & -39.77 \\
SNR & HESS J1614-518 & 243.56 & -51.82 \\
SNR & HESS J1457-593 & 223.7 & -59.07 \\
SNR & SNR G323.7-01.0 & 233.63 & -57.2 \\
SNR & HESS J1731-347 & 262.98 & -34.71 \\
SNR & Gamma Cygni & 305.27 & 40.52 \\
SNR & RCW 86 & 220.12 & -62.65 \\
SNR & HESS J1912+101 & 288.33 & 10.19 \\
SNR & HESS J1745-303 & 266.3 & -30.2 \\
SNR & Cassiopeia A & 350.85 & 58.81 \\ 
SNR & CTB 37A & 258.64 & -38.54 \\
\hline 
PWN & Vela X & 128.29 & -45.19 \\
PWN & Crab nebula & 83.63 & 22.01 \\
PWN & HESS J1708-443 & 257.0 & -44.3 \\
PWN & HESS J1825-137 & 276.55 & -13.58 \\
PWN & HESS J1632-478 & 248.01 & -47.87 \\
PWN & MSH 15-52 & 228.53 & -59.16 \\
PWN & HESS J1813-178 & 273.36 & -17.86 \\
PWN & HESS J1303-631 & 195.75 & -63.2 \\
PWN & HESS J1616-508 & 244.06 & -50.91 \\
PWN & HESS J1418-609 & 214.69 & -60.98  \\
PWN & HESS J1837-069 & 279.43 & -6.93  \\
PWN & HESS J1026-582 & 157.17 & -58.29 \\
\hline
UNID & MGRO J1908+06 & 286.91 & 6.32 \\
UNID & Westerlund 1 & 251.5 & -45.8 \\
UNID & HESS J1702-420 & 255.68 & -42.02 \\
UNID & 2HWC J1814-173 & 273.52 & -17.31 \\
UNID & HESS J1841-055 & 280.23 & -5.55 \\
UNID & 2HWC J1819-150 & 274.83 & -15.06 \\
UNID & HESS J1804-216 & 271.12 & -21.73 \\
UNID & HESS J1809-193 & 272.63 & -19.3 \\
UNID & HESS J1843-033 & 280.75 & -3.3 \\
UNID & TeV J2032+4130 & 307.93 & 41.51 \\
UNID & HESS J1708-410 & 257.10 & -41.09 \\
UNID & HESS J1857+026 & 284.30 & 2.67 \\
\hline
    \end{tabular}

    \label{tab:stacking_catalog}
\end{table}

\subsubsection*{Source List}

\begin{figure}
    \centering
    \includegraphics[width=\textwidth]{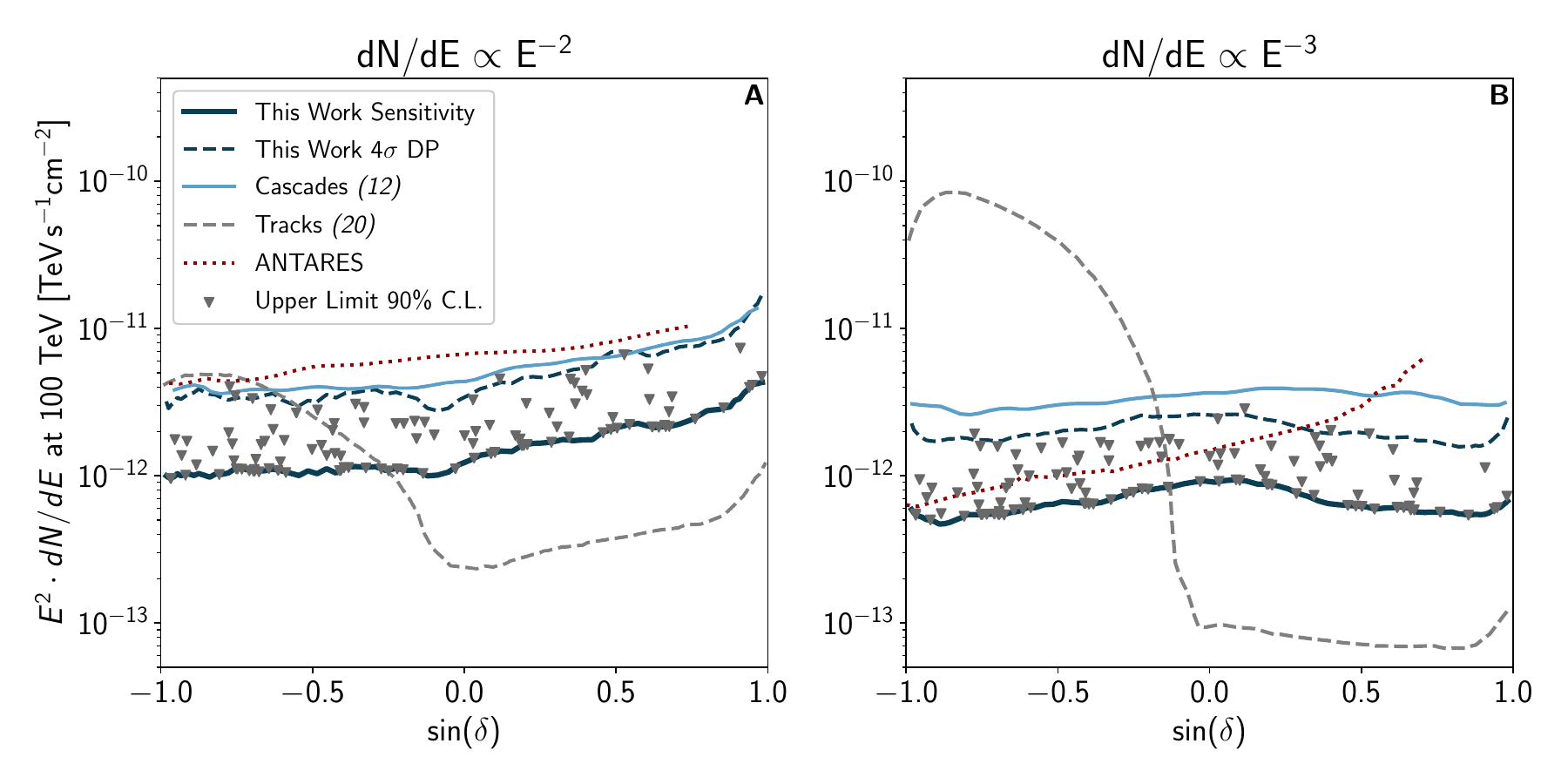}
    \caption{\textbf{Source list sensitivity and upper limits} Sensitivity to sources emitting an E$^{-2}$ spectrum (A) and E$^{-3}$ spectrum (B) for each data set. Individual sources in the source catalog are shown with their 90\% confidence level (CL) upper limits assuming an E$^{-2}$ (A) and E$^{-3}$ (B) emission spectra. ANTARES results are for E$^{-2}$~\cite{ICRC21_ANTARES_13yr}  and E$^{-3}$~\cite{ICRC19_ANTARES_11yr}  sensitivities. We also show previous results from IceCube tracks \cite{10yrtracks} and cascades \cite{mesecascades}.  Also shown in the 4$\sigma$ discovery potential (DP) for this work.  All results are consistent with background.}
    \label{fig:sens_ul}
\end{figure}

A search was performed using a list of 109 a priori selected positions in the sky.  The source list was constructed based on the GeV gamma-ray flux \cite{4FGL} catalog with previously described methods~\cite{10yrtracks} but optimized for the declination dependent sensitivity of this event selection. The list consists mainly of extragalactic objects. A point source likelihood test is performed at each position, and the results are shown in Table \ref{tab:source_list}. All results are consistent with background, and 90\% confidence flux upper limits are placed on the source flux $\Phi$ at $E_{\nu} = 100$~TeV assuming an $E^{-2}$ ($\Phi_2$) or $E^{-3}$ ($\Phi_3$) spectrum.  Upper limits for under-fluctuations are shown at the point source sensitivity by convention. The declination dependent sensitivity is shown in Figure \ref{fig:sens_ul} along with upper limits for each spectrum.  This work is an improvement to the sensitivity in the Southern Sky, however track-based analyses are still much more sensitive in the Northern Sky and individual point sources there are expected to emerge in track analyses before cascade analyses.

\newpage

\newpage
\setlength{\LTcapwidth}{\textwidth}

\begin{longtable}[t!]{cc|c|c|c|c|c|c|c}
\caption{\textbf{Source list search results.}  Sources in the source-list analysis, location~\cite{4FGL}, and pre-trial p-values.  Per-flavor 90\% flux upper limits are shown as E$^2  \frac{dN}{dE}$ at 100\,TeV in units of 10$^{-12}$ TeV\,cm$^{-2}$\,s$^{-1}$ for sources emitting following an E$^{-2}$ (UL $\Phi_2$) and  E$^{-3}$ (UL $\Phi_3$) spectrum. }\\
\hline
  & Source Name & $\alpha$ [$^\circ$] & $\delta$ [$^\circ$] & n$_s$ & $\gamma$ & p-value & UL $\Phi_2$ & UL $\Phi_3$ \\
 \hline
\endfirsthead
\hline
  & Source Name & $\alpha$ [$^\circ$] & $\delta$ [$^\circ$] & n$_s$ & $\gamma$ & p-value & UL $\Phi_2$ & UL $\Phi_3$ \\
 \hline
 \endhead

1 & G343.1-2.3 & 257.0 & -44.3  & 77.6 & 2.9  & 0.011 & \textless3.34 & \textless1.58 \\
2 & HESS J0835-455 & 128.3 & -45.2  & 0.2 & 2.0  & 0.719 & \textless0.59 & \textless0.28 \\
3 & PKS 0426-380 & 67.2 & -37.9  & 3.0 & 2.3  & 0.640 & \textless0.81 & \textless0.42 \\
4 & PKS 2155-304 & 329.7 & -30.2  & 61.5 & 4.0  & 0.249 & \textless1.52 & \textless1.02 \\
5 & Mkn 421 & 166.1 & 38.2  & 0.0 & -- & -- & \textless1.13 & \textless0.32 \\
6 & PKS 0537-441 & 84.7 & -44.1  & 1.1 & 1.8  & 0.576 & \textless0.89 & \textless0.45 \\
7 & PKS 0447-439 & 72.4 & -43.8  & 0.7 & 2.1  & 0.698 & \textless0.62 & \textless0.31 \\
8 & BL Lac & 330.7 & 42.3  & 7.6 & 1.6  & 0.340 & \textless2.73 & \textless0.77 \\
9 & PG 1553+113 & 238.9 & 11.2  & 0.0 & -- & -- & \textless0.68 & \textless0.35 \\
10 & TXS 0518+211 & 80.4 & 21.2  & 149.3 & 3.4  & 0.035 & \textless4.24 & \textless1.59 \\
11 & PKS 0235+164 & 39.7 & 16.6  & 0.0 & -- & -- & \textless0.67 & \textless0.31 \\
12 & PKS 1424+240 & 216.8 & 23.8  & 7.7 & 1.3  & 0.095 & \textless3.55 & \textless1.26 \\
13 & 3C 66A & 35.7 & 43.0  & 85.8 & 3.7  & 0.238 & \textless3.44 & \textless0.89 \\
14 & TXS 0506+056 & 77.4 & 5.7  & 0.0 & -- & -- & \textless0.61 & \textless0.39 \\
15 & AP Librae & 229.4 & -24.4  & 0.3 & 1.0  & 0.725 & \textless0.61 & \textless0.34 \\
16 & S5 0716+71 & 110.5 & 71.3  & 1.9 & 1.0  & 0.532 & \textless3.94 & \textless0.58 \\
17 & B2 1215+30 & 184.5 & 30.1  & 5.2 & 2.2  & 0.573 & \textless1.80 & \textless0.53 \\
18 & MH 2136-428 & 324.9 & -42.6  & 0.0 & -- & -- & \textless0.53 & \textless0.27 \\
19 & PKS 2233-148 & 339.1 & -14.6  & 25.8 & 4.0  & 0.584 & \textless0.93 & \textless0.65 \\
20 & Mkn 501 & 253.5 & 39.8  & 4.1 & 1.0  & 0.514 & \textless2.09 & \textless0.60 \\
21 & PMN J1603-4904 & 241.0 & -49.1  & 84.5 & 3.1  & 0.010 & \textless3.49 & \textless1.59 \\
22 & S2 0109+22 & 18.0 & 22.8  & 147.8 & 4.0  & 0.077 & \textless3.75 & \textless1.31 \\
23 & PKS 0301-243 & 45.9 & -24.1  & 44.3 & 3.6  & 0.405 & \textless1.37 & \textless0.76 \\
24 & 4C +01.28 & 164.6 & 1.6  & 16.1 & 1.2  & 0.025 & \textless3.27 & \textless2.42 \\
25 & PKS 0700-661 & 105.1 & -66.2  & 28.8 & 2.9  & 0.239 & \textless1.72 & \textless0.82 \\
26 & TXS 0628-240 & 97.7 & -24.1  & 0.9 & 3.0  & 0.746 & \textless0.55 & \textless0.31 \\
27 & PKS 0823-223 & 126.5 & -22.5  & 2.0 & 2.0  & 0.553 & \textless1.01 & \textless0.58 \\
28 & PKS 0735+17 & 114.5 & 17.7  & 63.6 & 4.0  & 0.376 & \textless2.16 & \textless0.91 \\
29 & PMN J1329-5608 & 202.3 & -56.1  & 40.3 & 4.0  & 0.276 & \textless1.48 & \textless0.77 \\
30 & PMN J0531-4827 & 83.0 & -48.5  & 0.0 & -- & -- & \textless0.55 & \textless0.26 \\
31 & MG1 J021114+1051 & 32.8 & 10.9  & 28.7 & 4.0  & 0.622 & \textless1.22 & \textless0.66 \\
32 & PKS 1440-389 & 221.0 & -39.1  & 11.0 & 1.0  & 0.175 & \textless2.07 & \textless1.10 \\
33 & OT 081 & 267.9 & 9.6  & 35.2 & 2.8  & 0.356 & \textless1.88 & \textless1.10 \\
34 & OJ 287 & 133.7 & 20.1  & 28.6 & 2.7  & 0.473 & \textless1.84 & \textless0.74 \\
35 & PKS 1101-536 & 166.0 & -54.0  & 0.0 & -- & -- & \textless0.46 & \textless0.23 \\
36 & TXS 0141+268 & 26.2 & 27.1  & 1.4 & 3.2  & 0.706 & \textless1.17 & \textless0.37 \\
37 & 1H 1013+498 & 153.8 & 49.4  & 4.2 & 2.6  & 0.676 & \textless1.65 & \textless0.36 \\
38 & PKS 0048-09 & 12.7 & -9.5  & 115.2 & 3.6  & 0.089 & \textless2.35 & \textless1.68 \\
39 & PMN J1650-5044 & 252.6 & -50.8  & 88.1 & 2.9  & 0.002 & \textless4.02 & \textless1.93 \\
40 & PKS 0118-272 & 20.1 & -27.0  & 44.8 & 3.9  & 0.378 & \textless1.38 & \textless0.82 \\
41 & 1H 1914-194 & 289.4 & -19.4  & 26.9 & 2.5  & 0.152 & \textless2.30 & \textless1.26 \\
42 & PKS 0332-403 & 53.6 & -40.1  & 9.8 & 2.4  & 0.481 & \textless1.13 & \textless0.59 \\
43 & OJ 014 & 122.9 & 1.8  & 2.6 & 4.0  & 0.771 & \textless0.57 & \textless0.39 \\
44 & PMN J1918-4111 & 289.6 & -41.2  & 54.6 & 3.7  & 0.256 & \textless1.72 & \textless0.89 \\
45 & PKS 1936-623 & 295.3 & -62.2  & 26.6 & 4.0  & 0.420 & \textless1.19 & \textless0.55 \\
46 & 1H 1720+117 & 261.3 & 11.9  & 6.8 & 2.7  & 0.712 & \textless0.91 & \textless0.49 \\
47 & PMN J1610-6649 & 242.7 & -66.8  & 10.1 & 2.9  & 0.588 & \textless0.84 & \textless0.41 \\
48 & PMN J0334-3725 & 53.6 & -37.4  & 10.7 & 2.4  & 0.456 & \textless1.25 & \textless0.65 \\
49 & TXS 1714-336 & 259.4 & -33.7  & 67.7 & 2.9  & 0.028 & \textless2.68 & \textless1.54 \\
50 & PKS 2005-489 & 302.4 & -48.8  & 0.0 & -- & -- & \textless0.54 & \textless0.26 \\
51 & PKS B1056-113 & 164.8 & -11.6  & 19.6 & 2.1  & 0.096 & \textless2.27 & \textless1.66 \\
52 & RGB J2243+203 & 341.0 & 20.4  & 163.1 & 4.0  & 0.022 & \textless4.53 & \textless1.82 \\
53 & 1ES 1959+650 & 300.0 & 65.1  & 150.0 & 4.0  & 0.071 & \textless7.30 & \textless1.14 \\
54 & S4 0814+42 & 124.6 & 42.4  & 15.4 & 4.0  & 0.568 & \textless1.74 & \textless0.49 \\
55 & KUV 00311-1938 & 8.4 & -19.4  & 109.8 & 3.6  & 0.063 & \textless2.91 & \textless1.61 \\
56 & PMN J2250-2806 & 342.7 & -28.1  & 59.1 & 4.0  & 0.244 & \textless1.62 & \textless1.05 \\
57 & 1RXS J130421.2-435308 & 196.1 & -43.9  & 7.1 & 2.3  & 0.473 & \textless1.11 & \textless0.57 \\
58 & PMN J0810-7530 & 122.8 & -75.5  & 16.5 & 4.0  & 0.598 & \textless0.79 & \textless0.44 \\
59 & 3C 454.3 & 343.5 & 16.2  & 106.2 & 4.0  & 0.188 & \textless2.67 & \textless1.25 \\
60 & PKS 1424-41 & 217.0 & -42.1  & 10.1 & 1.0  & 0.280 & \textless1.63 & \textless0.83 \\
61 & 3C 279 & 194.0 & -5.8  & 10.1 & 1.0  & 0.120 & \textless1.90 & \textless1.63 \\
62 & CTA 102 & 338.2 & 11.7  & 129.6 & 4.0  & 0.102 & \textless3.09 & \textless1.60 \\
63 & PKS 1510-089 & 228.2 & -9.1  & 4.8 & 1.0  & 0.212 & \textless1.79 & \textless1.34 \\
64 & PKS 0454-234 & 74.3 & -23.4  & 0.0 & -- & -- & \textless0.55 & \textless0.30 \\
65 & PKS 1502+106 & 226.1 & 10.5  & 8.7 & 1.0  & 0.422 & \textless1.79 & \textless0.99 \\
66 & PKS 1830-211 & 278.4 & -21.1  & 70.4 & 2.8  & 0.033 & \textless3.07 & \textless1.68 \\
67 & PKS 2326-502 & 352.3 & -49.9  & 6.6 & 2.1  & 0.265 & \textless1.66 & \textless0.84 \\
68 & PKS 0727-11 & 112.6 & -11.7  & 8.2 & 2.6  & 0.598 & \textless0.83 & \textless0.65 \\
69 & 4C +21.35 & 186.2 & 21.4  & 11.4 & 1.5  & 0.162 & \textless3.09 & \textless1.15 \\
70 & PMN J2345-1555 & 356.3 & -15.9  & 11.9 & 3.3  & 0.646 & \textless0.78 & \textless0.53 \\
71 & 4C +01.02 & 17.2 & 1.6  & 74.8 & 3.8  & 0.336 & \textless1.66 & \textless1.18 \\
72 & PKS 2023-07 & 306.4 & -7.6  & 122.8 & 3.5  & 0.060 & \textless2.32 & \textless1.76 \\
73 & Ton 599 & 179.9 & 29.2  & 14.5 & 2.4  & 0.405 & \textless2.49 & \textless0.74 \\
74 & 4C +38.41 & 248.8 & 38.1  & 3.5 & 1.0  & 0.573 & \textless1.84 & \textless0.54 \\
75 & PKS 1244-255 & 191.7 & -25.8  & 3.9 & 1.0  & 0.128 & \textless2.04 & \textless1.28 \\
76 & B2 1520+31 & 230.5 & 31.7  & 9.3 & 1.0  & 0.013 & \textless6.59 & \textless1.93 \\
77 & PKS 1124-186 & 171.8 & -19.0  & 0.0 & -- & -- & \textless0.56 & \textless0.33 \\
78 & 4C +28.07 & 39.5 & 28.8  & 0.0 & -- & -- & \textless1.15 & \textless0.36 \\
79 & PKS 1730-13 & 263.3 & -13.1  & 51.2 & 2.7  & 0.128 & \textless2.27 & \textless1.59 \\
80 & PKS 0805-07 & 122.1 & -7.9  & 2.4 & 2.3  & 0.676 & \textless0.69 & \textless0.53 \\
81 & PKS 0208-512 & 32.7 & -51.0  & 4.9 & 1.8  & 0.144 & \textless1.97 & \textless1.03 \\
82 & PKS 0336-01 & 54.9 & -1.8  & 11.0 & 4.0  & 0.729 & \textless0.62 & \textless0.47 \\
83 & PKS 0502+049 & 76.3 & 5.0  & 0.0 & -- & -- & \textless0.61 & \textless0.39 \\
84 & PKS 0402-362 & 61.0 & -36.1  & 2.2 & 2.2  & 0.672 & \textless0.71 & \textless0.38 \\
85 & PMN J1802-3940 & 270.7 & -39.7  & 76.7 & 3.1  & 0.043 & \textless2.81 & \textless1.39 \\
86 & PKS 1622-253 & 246.4 & -25.5  & 36.6 & 2.6  & 0.089 & \textless2.26 & \textless1.36 \\
87 & PKS 2052-47 & 314.1 & -47.2  & 0.0 & -- & -- & \textless0.55 & \textless0.26 \\
88 & 3C 273 & 187.3 & 2.0  & 11.3 & 1.0  & 0.227 & \textless2.01 & \textless1.41 \\
89 & PKS 2142-75 & 326.8 & -75.6  & 2.6 & 1.0  & 0.744 & \textless0.49 & \textless0.26 \\
90 & MG1 J123931+0443 & 189.9 & 4.7  & 9.3 & 1.0  & 0.217 & \textless2.21 & \textless1.42 \\
91 & MG2 J201534+3710 & 303.9 & 37.2  & 113.6 & 3.1  & 0.039 & \textless5.29 & \textless1.51 \\
92 & Galactic Centre & 266.4 & -29.0  & 58.7 & 2.7  & 0.022 & \textless2.79 & \textless1.67 \\
93 & PKS 2247-131 & 342.5 & -12.8  & 7.3 & 4.0  & 0.709 & \textless0.62 & \textless0.46 \\
94 & NGC 1275 & 50.0 & 41.5  & 29.3 & 3.3  & 0.529 & \textless2.01 & \textless0.57 \\
95 & PKS 0521-36 & 80.7 & -36.5  & 5.7 & 2.0  & 0.237 & \textless1.75 & \textless1.00 \\
96 & Cen A & 201.0 & -43.5  & 9.4 & 2.4  & 0.398 & \textless1.30 & \textless0.66 \\
97 & LMC & 80.0 & -68.8  & 31.5 & 4.0  & 0.327 & \textless1.38 & \textless0.71 \\
98 & SMC & 14.5 & -72.8  & 43.1 & 1.0  & 0.174 & \textless1.77 & \textless0.94 \\
99 & NGC 4945 & 196.4 & -49.5  & 7.3 & 2.4  & 0.425 & \textless1.28 & \textless0.63 \\
100 & NGC 253 & 11.9 & -25.3  & 51.4 & 3.8  & 0.328 & \textless1.43 & \textless0.88 \\
101 & NGC 1068 & 40.7 & -0.0  & 87.1 & 4.0  & 0.252 & \textless1.88 & \textless1.36 \\
102 & M 82 & 148.9 & 69.7  & 22.9 & 2.9  & 0.564 & \textless3.44 & \textless0.53 \\
103 & Arp 220 & 233.7 & 23.5  & 15.2 & 1.0  & 0.009 & \textless5.18 & \textless2.03 \\
104 & M 31 & 10.8 & 41.2  & 18.2 & 2.9  & 0.487 & \textless2.21 & \textless0.62 \\
105 & NGC 3424 & 162.9 & 32.9  & 0.0 & -- & -- & \textless1.19 & \textless0.34 \\
106 & IC 678 & 168.6 & 6.6  & 19.2 & 1.8  & 0.005 & \textless4.55 & \textless2.85 \\
107 & NGC 5380 & 209.3 & 37.5  & 10.6 & 1.0  & 0.249 & \textless3.30 & \textless0.93 \\
108 & Arp 299 & 172.1 & 58.5  & 0.0 & -- & -- & \textless1.25 & \textless0.23 \\
109 & NGC 2146 & 94.5 & 78.3  & 4.2 & 1.0  & 0.445 & \textless4.71 & \textless0.73 \\
\hline


\label{tab:source_list}
\end{longtable}

\FloatBarrier



\clearpage


\end{document}